\documentclass[11pt, a4paper]{article}
\usepackage[english]{babel}
\usepackage[doublespacing]{setspace}
\usepackage{array}
\usepackage[flushleft]{threeparttable}
\usepackage{longtable}
\usepackage{appendix}
\usepackage{float}
\usepackage{amssymb}
\usepackage{marvosym}
\usepackage{ragged2e}
\usepackage{dsfont}
\usepackage{lmodern}
\usepackage{mathdots}
\usepackage{amsmath}
\usepackage{amsthm}
\usepackage{mathtools}
\usepackage{upgreek}
\usepackage{graphicx}
\usepackage{tikz}
\usepackage{xcolor}
\usepackage{marginnote}
\usepackage[round, sort]{natbib}
\usepackage{subcaption}
\usepackage{changepage}
\usepackage{multirow}
\usepackage{authblk}
\usepackage{booktabs}
\usepackage{braket}
\usepackage{enumitem}
\usepackage{adjustbox}
\usepackage{natbib}
\usepackage[sc]{mathpazo}
\usepackage[font = small, labelfont = bf]{caption}
\usepackage[a4paper, bindingoffset = 0.2in, %
left = 0.8in, right = 0.9in, top = 1in, bottom = 1in, %
footskip = .25in]{geometry}

\usepackage{algorithm}

\usepackage{xcolor}

\DeclareMathAlphabet{\mathpzc}{OT1}{pzc}{m}{it}

\DeclareMathOperator{\E}{\mathbb{E}}
\DeclareMathOperator{\Cov}{\mathbb{C}ov}

\DeclareMathOperator{\Cum}{\mathbb{C}um}

\DeclareMathOperator{\Id}{\boldsymbol{I}}

\DeclareMathOperator{\rank}{\text{rank}}
\DeclareMathOperator{\vect}{\text{vec}}

\DeclarePairedDelimiter{\floor}{\lfloor}{\rfloor}

\newcommand{\bs}[1]{\boldsymbol{#1}}

\newcommand*{\mbb}{\mathbb}
\newcommand*{\mb}{\mathbf}
\newcommand{\norm}[1]{\left\|{#1}\right\|}
\newcommand{\abs}[1]{\left|{#1}\right|}
\newcommand{\longbar}[1]{\overline{#1}}

\numberwithin{equation}{section}
\numberwithin{figure}{section}
\usepackage[xcolor, leftbars]{changebar}
\definecolor{darkblue}{rgb}{0.01, 0.2, 1.0}
\usepackage[unicode, bookmarks = FALSE,colorlinks=true,allcolors=darkblue ]{hyperref}
\def\equationautorefname~#1\null{%
	Equation~(#1)\null
}
\let\oldFootnote\footnote
\newcommand\nextToken\relax

\renewcommand\footnote[1]{%
	\oldFootnote{#1}\futurelet\nextToken\isFootnote}

\newcommand\isFootnote{%
	\ifx\footnote\nextToken\textsuperscript{,}\fi}

\usepackage{array}
\newcolumntype{L}[1]{>{\raggedright\let\newline\\\arraybackslash\hspace{0pt}}m{#1}}
\newcolumntype{C}[1]{>{\centering\let\newline\\\arraybackslash\hspace{0pt}}m{#1}}
\newcolumntype{R}[1]{>{\raggedleft\let\newline\\\arraybackslash\hspace{0pt}}m{#1}}
\newcommand*{\defeq}{\mathrel{\vcenter{\baselineskip0.5ex \lineskiplimit0pt
			\hbox{\scriptsize.}\hbox{\scriptsize.}}}%
	=}

\newtheorem{assumption}{Assumption}[section]
\newtheorem{proposition}{Proposition}[section]
\newtheorem{corollary}{Corollary}[section]

\title{\textbf{Robust Estimation of the non-Gaussian Dimension in Structural Linear Models}}
\author{Miguel Angel Cabello\thanks{Email: micabell@eco.uc3m.es. I thank to my doctoral supervisor, Carlos Velasco, as well as Miguel Angel Delgado, Juan José Dolado, Juan Carlos Escanciano and Jesús Gonzalo, for their valuable comments to improve this paper. I am responsible of all mistakes and omissions. Financial support by MCIN/BES-2017-081997,  MCIN/AEI /10.13039/501100011033-CEX2021-001181-M, and Comunidad de Madrid, grants EPUC3M11 (V PRICIT) and H2019/HUM-5891, is gratefully acknowledged.} \\ Universidad Carlos III de Madrid and Central Bank of Peru}

\begin{document}
	\maketitle
	
	\begin{abstract}
		Statistical identification of possibly non-fundamental SVARMA models requires structural errors: (i) to be an i.i.d process, (ii) to be mutually independent across components, and (iii) each of them must be non-Gaussian distributed. Hence, provided the first two requisites, it is crucial to evaluate the non-Gaussian identification condition. We address this problem by relating the non-Gaussian dimension of structural errors vector to the rank of a matrix built from the higher-order spectrum of reduced-form errors. This makes our proposal robust to the roots location of the lag polynomials, and generalizes the current procedures designed for the restricted case of a causal structural VAR model. Simulation exercises show that our procedure satisfactorily estimates the number of non-Gaussian components.  
	\end{abstract}
	\noindent%
	
	\section{Introduction}
	
	Central to assessing and quantifying the effects of economic shocks over real activity variables and other macroeconomic outcomes is the identification of such disturbances. For achieving this, it is usual to employ structural vector auto-regressive models and to apply a particular identification strategy. Most of these strategies resort on imposing external restrictions, which may come from a theoretical model, a particular data frequency or by exploiting granular institutional information (see \citet{sims1980macroeconomics,blanchard1989dynamic,bernanke1998measuring,blanchard2002empirical,gali1992well} as examples of imposing zero, linear and non-linear identification restrictions, economically motivated, over VAR parameters; \citet{uhlig2005effects,canova2007price,arias2018inference} for agnostic economically motivated restrictions; \citet{romer1989does,romer2010macroeconomic} for narrative identification strategy; and \citet{mertens2013dynamic,mertens2014reconciliation,stock2018identification} for instrumental variables identification approach).
	
	
	Alternatively, the statistical identification strategy (hereafter SIS) of structural VAR models has appeared on the scene as part of a \textit{data-driven} identification approach and has gained relevance, because this strategy permits to identify a SVAR model without employing external information and makes feasible to evaluate any economically motivated restriction. The seminal work of \citet{comon1994independent} states that a linear transformation of a vector whose components are mutually independent and with at most one Gaussian distributed element is identified up to signed permutation (see \citet{lanne2017identification,maxand2020identification,guay2021identification} for a direct application within the context of a causal SVAR model). For general linear, stationary processes, the non-Gaussian requirement needs to be satisfied by each of the structural shocks. Particularly, \citet{chan2006note} show that identification of a linear, possibly non-causal, stationary process needs that structural shocks vector must be fully independent\footnote{Being \textit{fully independent} implies that a shocks is serial and contemporaneously independent.} and each shock must have non-zero and finite third and fourth order cumulants, respectively. This result allows is the cornerstone to statistically identify possibly non-fundamental SVARMA models\footnote{Non-fundamental behavior is equivalent to AR and MA lag polynomials have some roots inside the unit circle} (see \citet{lanne2010structural,gourieroux2020identification} for applications with likelihood methods and \citet{velasco2022identification} for an equivalent identification result through higher-order cumulant conditions). 
	
	In contrast to other identification strategies, the SIS permits  to evaluate the validity of its identification assumptions. For instance, \citet{hong1999hypothesis} develops a procedure for testing the serial independence assumption. Also, \citet{amengual2022moment} provide a framework for evaluating the contemporaneous independence and the non-Gaussianity of structural shocks. In this paper, we focus on developing a procedure for evaluating the non-Gaussian identification condition, since even though serial and mutual independence hold, non-Gaussian behavior of structural shocks is decisive for making feasible application of the SIS. Besides, as it is detailed below, our proposal seeks to be implemented without requiring estimation of structural model. Thus, the procedure can be performed before applying the SIS.
	
	There is a voluminous literature coping with the problem of determining whether a random variable (or vector) is  Gaussian distributed or not. Some procedures are based on the empirical distribution of data (see \citet{kolmogorov1933sulla,smirnov1948table,massey1951kolmogorov,shapiro1965analysis}); other approaches employ the characteristic function (see \citet{hall1983test,epps1983test}); while others exploit third and fourth order centered moments (see \citet{d1971omnibus,bera1982model,lobato2004simple}). These procedures only assess the hypothesis of joint Gaussianity, which in case of no rejection, would make infeasible to apply the SIS to identify structural parameters in a SVARMA model. Nonetheless, if joint Gaussianity were rejected, this would support only the existence of at least one non-Gaussian component in the system, and this conclusion would not be sufficient for justifying the application of the SIS.
	
	The literature of estimating the non-Gaussian dimension or the number of non-Gaussian distributed shocks inside a random vector is less prolific. \citet{nordhausen2017asymptotic} propose both asymptotic and bootstrap tests to estimate the non-Gaussian dimension of an unobservable random vector by analyzing the number of non-zero eigenvalues of a convenient transformation of the scatter matrix\footnote{The scatter matrix contains fourth-order moments of a vector.} constructed using information from an observable random vector, which is an affine transformation of the unobservable one.\footnote{Let $\mb{x}$ and $\bs{\varepsilon}$ be an observable and unobservable $d$-dimensional random vectors, respectively. An affine transformation is define as $\mb{x} = \bs{b}_{0} + \bs{B}\bs{\varepsilon}$, with $\bs{b}_{0}$ possibly non-zero vector and $\bs{B}$ time-invariant, full rank, square matrix.} They find the null eigenvalues of the scatter matrix is equal to the number of Gaussian distributed components in the vector of unobservable shocks.\footnote{To be completely precise, there exists $r$ eigenvalues equal to $d+2$, but \citet{nordhausen2017asymptotic} employ a normalized version of the scatter matrix.} Within the context of a causal SVAR models, \citet{maxand2020identification} applies \cite{nordhausen2017asymptotic} approach directly to estimated structural shocks, because according to her, the non-Gaussian block remains identified even though there is more than one Gaussian shock in the system. Following a similar approach, \citet{amengual2022moment} estimates structural VAR model by pseudo MLE and applies Normality test to each estimated structural error. Alternatively, \citet{guay2021identification} approach does not require to estimate the structural model, but only the reduced-form VAR model. He relates the non-Gaussian dimension to the rank of the matrix version of fourth order cumulants of the reduced-form errors.\footnote{When third and fourth order moments exist, the third order cumulant is the same as the asymmetry coefficient, while the fourth order cumulant represents the excess of kurtosis. Gaussian distribution is the only one with all cumulants of third or higher order equal to zero. See \citet{marcinkiewicz1939propriete} for technical details.} Guay's approach reduces the original task to a matrix rank estimation problem, thus he directly applies \citet{robin2000tests} sequential procedure.
	
	These approaches are helpful but restrictive for some reasons. First, they consider only the particular case of having a causal structural VAR model as the true, underlying data generating process. However, many structural macroeconomic models can be represented and fitted more accurately by an SVARMA model rather than an SVAR one (see \citet{ravenna2007vector,fernandez2007abcs}). Second, as discussed in \citet{velasco2022identification,gourieroux2020identification} and the references therein, the fundamentalness assumption is necessary for avoiding the dynamic identification problem -i.e., for any stationary linear process, its fundamental and non-fundamental representations are observationally equivalent when only second-order information is exploited. And, as surveyed by \citet{alessi2011non}, non-fundamentalness may be consistent with many macroeconomic models and data features. 
	
	Consequently, this article attempts to fill this gap in the literature and proposes a procedure for determining the non-Gaussian dimension in a SVARMA model, without requiring prior knowledge about the roots location of the lag polynomials. Nonetheless, this is a quite challenging task, since the reduced-form (hereafter RF) errors from a fundamental VARMA approximation -or RF-VARMA model-, when the roots location of lag polynomials is unknown, might not be a simple static rotation of structural shocks, but a dynamic filter of them. This characteristic makes invalid \citet{guay2021identification}, \citet{maxand2020identification} and \citet{amengual2022moment} approaches, because the rank or the number of non-zero eigenvalues of a matrix constructed from contemporaneous fourth-order cumulants of the RF errors are not related one-to-one to the non-Gaussian dimension, unless extreme assumptions are made. In contrast, we exploit third or fourth-order cumulant spectrums of RF errors for constructing a matrix whose rank unveils the non-Gaussian dimension in the vector of structural shocks. Our work can be seen as the extension of \citet{lobato2004simple} approach to a multivariate context and the generalization of \citet{guay2021identification} and \cite{amengual2022moment} works to situations in which roots location is unknown. 	
	
	\textcolor{red}{CHANGE THIS PART!!}
	For estimating the rank of a matrix, there exist several strategies in the literature. We follow \citet{kleibergen2006generalized} (hereafter KP) approach. The KP  statistic is built from the singular value decomposition of the matrix of interest. The asymptotic distribution of the statistic is a standard chi-square whose degrees of freedom change depending on the null hypothesis. Unlike KP work, our context changes the asymptotic distribution of the test statistic for some particular rank values, specifically under joint Gaussianity. For other null hypotheses, the asymptotic distribution is chi-square, but the degrees of freedom are generally unknown. Thus, we propose a bootstrap strategy. This path implies another challenge: to impose the null hypothesis in the resampled data. Montecarlo exercises were performed to analyze the size and power of the bootstrap test; these show that our strategy estimates satisfactorily the non-Gaussian dimension. We apply our procedure to two well-known macroeconomic datasets. Our proposal detects a skewed structural shock in the system described by \citet{blanchard1989dynamic} (hereafter BQ). Using \citet{blanchard2002empirical} (hereafter BP), our approach detects at least two skewed and non-mesokurtic structural errors, unlike \citet{guay2021identification} whose procedure only could detect one non-mesokurtic structural shock. In the case of imposing roots outside the unit circle, this latter result implies that the SIS can be applied to BQ or BP datasets.
	
	The remainder of this paper is structured as follows: Section 2 describes our time series model and states the main assumptions. Section 3 shows the connection between the number of non-Gaussian structural shocks and the rank of a matrix constructed from third and fourth order spectrum of reduced-form errors. In Section 4, the test procedure and estimation are detailed. Section 5 shows simulation results and the empirical application. We conclude in Section 6. 
	
	\section{Model and Assumptions}
	\label{sec:Basics}
	Let $\bs{y}_{t}$ be a $d$-dimensional stationary, zero-mean process generated by a structural vector autoregression moving-average (SVARMA) model\footnote{See \citet{gourieroux2020identification,velasco2022identification,mainassara2011estimating} for alternative representations}
	\begin{equation}
		\label{eq:Eq1}
		\bs{\Phi}(L)\bs{y}_{t} = \bs{\Theta}(L)\bs{B}\bs{\varepsilon}_{t},
	\end{equation}
	where $\bs{\Phi}(L)=\Id_{d}-\sum_{j=1}^{p}{\bs{\Phi}_{j}L^{j}}$, $\bs{\Theta}(L)=\Id_{d}+\sum_{j=1}^{q}{\bs{\Theta}_{j}L^{j}}$. $L$ is the lag (back-shift) operator, i.e. $L^{j}\bs{y}_{t}=\bs{y}_{t-j}$ for any $j\in\mathbb{Z}$. $\bs{B}$ is a time-invariant, full-rank, squared matrix, also known as the contemporaneous effects matrix. $\bs{\varepsilon}_{t}$ is a $d$-dimensional vector, representing the structural shocks.\footnote{These shocks have economic interpretation.}
	
	The $\bs{\Phi}(z)$ and $\bs{\Theta}(z)$ represents the autoregressive (AR) and moving-average (MA) polynomials, respectively. These hold
	\begin{align}
		\label{eq:Eq2}
		\det\left(\bs{\Phi}(z)\right)\det\left(\bs{\Theta}(z)\right) & \neq{ 0},\quad \forall\; z\in{\mathbb{T}=\left\{ z\in\mathbb{C}\;|\;|z|=1 \right\}}.
	\end{align}
	Condition (\ref{eq:Eq2}) rules out only the presence of unit roots in both AR and MA polynomials, otherwise a stationary solution to equation (\ref{eq:Eq1}) does not exist. Unlike current approaches in the literature, this requirement does not impose an exact location of AR and MA polynomials roots\footnote{Compares this to alternative works such as \citet{maxand2020identification,guay2021identification,amengual2022moment}, where they assume that condition in (\ref{eq:Eq2}) holds for any $z\in\mathbb{T}_{+}=\left\{ z\in\mathbb{C}\;|\;|z|\leq1 \right\}$.}, implying that the model in (\ref{eq:Eq1}) allows for possibly non-fundamental behavior. 
	
	The structural shocks, $\bs{\varepsilon}_{t}$, have a specific behavior summarized by the following assumption.
	\begin{assumption}\mbox{}
		\label{as:Assumption1}
		\begin{enumerate}
			\item[(i)] $\bs{\varepsilon}_{t}$ is an independent, identically distributed (i.i.d.) process;
			\item[(ii)] $\bs{\varepsilon}_{t}$ has mutually independent components;
			\item[(iii)] there exists $0\leq d_{ng}\leq d$ non-Gaussian distributed shocks in the vector $\bs{\varepsilon}_{t}$;
			\item[(iv)] $\E[\bs{\varepsilon}_{t}]=\bs{0}$,  $\E[\bs{\varepsilon}_{t}\bs{\varepsilon}_{t}^{\prime}]=\Id_{d}$ and  $\E\left(\norm{\bs{\varepsilon}_{t}}^{8}\right)<\infty$ for any $t\in\mathbb{Z}$.
		\end{enumerate}
	\end{assumption}
	Assumption (\ref{as:Assumption1}.(i)) is standard in the literature\footnote{See for instance \citet{lanne2010structural,lanne2017identification,gourieroux2020identification,velasco2022identification,guay2021identification}.} and makes the analysis simpler. However, it is a restrictive requisite since it implies structural shocks do not exhibit any linear or nonlinear serial dependence\footnote{This condition rules out some characteristics that may be relevant in the empirical analysis of macroeconomic outcomes such as conditional heteroskedasticity.}. Assumption (\ref{as:Assumption1}.(ii)) is similar to the one stated in \citet{comon1994independent}. It is the basis of \textit{Independent Component Analysis} (hereafter ICA). This requirement is not as restricting as it seems, e.g., it is standard in macroeconomics to assume that productivity shock is independent of monetary or fiscal policy shocks.
	
	Assumption (\ref{as:Assumption1}.(iii)) does not impose a particular number of non-Gaussian shocks, because this will be determined empirically. This implies the structural model cannot be estimated, since it is unidentified. Besides, if $d_{ng}$ is the number of non-Gaussian components in $\bs{\varepsilon}_{t}$, the rest $d-d_{ng}$ structural errors are Gaussian distributed. Finally, Assumption (\ref{as:Assumption1}.(iv)) imposes structural errors are centered and standardized with finite moments up to eighth order. This last requirement is necessary, because we will characterize non-Gaussian behavior using third and fourth order information.
	
	The vector of structural parameters of the model in equation (\ref{eq:Eq1}) can be represented by $\bs{\vartheta}_{0}$, a $K$-dimensional column vector. We split $\bs{\vartheta}_{0}$ into two blocks: one related to dynamic behavior ($\bs{\vartheta}_{0,1}$) and another that governs the static component ($\bs{\vartheta}_{0,2}$). Hence, the structural model can be written as $\bs{\Phi}(L,\bs{\vartheta}_{0,1})\bs{y}_{t} = \bs{\Theta}(L,\bs{\vartheta}_{0,1})\bs{B}(\bs{\vartheta}_{0,2})\bs{\varepsilon}_{t}$. The moving-average representation of the observable variables is $\bs{y}_{t}=\bs{\Psi}(L,\bs{\vartheta}_{0,1})\bs{B}(\bs{\vartheta}_{0,2})\bs{\varepsilon}_{t}$, where $\bs{\Psi}(L,\bs{\vartheta}_{0,1})\defeq\bs{\Phi}^{-1}(L,\bs{\vartheta}_{0,1})\bs{\Theta}(L,\bs{\vartheta}_{0,1})$ is a possibly non-causal filter.\footnote{A non-causal filter is of the form $\bs{\Psi}(L)=\sum_{j=-\infty}^{\infty}{\bs{\Psi}_{j}L^{j}}$ with non-zero $\bs{\Psi}_{j}$ for some $j<0$.} The vector of structural parameters, $\bs{\vartheta}_{0}$, is not identified, because Assumption (\ref{as:Assumption1}) does not assure the non-Gaussian behavior of the structural shocks. Identification assuming that all roots of AR and MA polynomials are outside unit circle requires $d_{ng}\geq{d-1}$ (\citet{comon1994independent}); while, when roots location is unknown and only unit roots are discarded $d_{ng}=d$ is needed (\citet{chan2006note,velasco2020identification}).
	
	Instead of working with structural estimates, which requires identification of $\bs{\vartheta}_{0}$, in this research we propose a procedure that exploits only reduced-form estimates, for which structural parameters identification is unnecessary. According to Wold's Decomposition Theorem (hereafter WDT), any stationary process $\bs{y}_{t}$ can be represented as a square summable, infinite, causal moving-average of serially uncorrelated errors (see \citet{anderson2011statistical} for more details). This means that $\bs{y}_{t}=\tilde{\bs{\Psi}}(L)\bs{u}_{t}$ with $\tilde{\bs{\Psi}}(L)=\sum_{j=0}^{\infty}{\tilde{\bs{\Psi}}_{j}L^{j}}$ such that $\sum_{j=0}^{\infty}{\norm{\tilde{\bs{\Psi}}_{j}}^{2} }<\infty$, $\tilde{\bs{\Psi}}_{0}=\bs{I}_{d}$ and $\Cov\left( \bs{u}_{t}, \bs{u}_{t-k} \right)=\bs{0}$ for all $\abs{k}\geq1$. Without loss of generality, we can assume that $\tilde{\bs{\Psi}}(L)=\tilde{\bs{\Phi}}^{-1}(L)\tilde{\bs{\Theta}}(L)$ with $\det\left(\tilde{\bs{\Phi}}(z)\right)\det\left(\tilde{\bs{\Theta}}(z)\right)\neq{0}$ for all $\abs{z}\leq1$. Furthermore, let $\bs{\vartheta}_{f}$ denotes the parameters vector of the fundamental approximation of $\bs{y}_{t}$, thus $\bs{y}_{t}=\bs{\Psi}(L,\bs{\vartheta}_{f})\bs{u}_{t}=\bs{\Phi}^{-1}(L,\bs{\vartheta}_{f})\bs{\Theta}(L,\bs{\vartheta}_{f})\bs{u}_{t}$. From this and equation (\ref{eq:Eq1}), the serially uncorrelated RF errors, $\bs{u}_{t}$, are equal to
	\begin{equation}
		\label{eq:ReducedFormErrors}
		\bs{u}_{t}\defeq\bs{\Psi}^{-1}(L,\bs{\vartheta}_{f})\bs{y}_{t} = \bs{\Psi}^{-1}(L,\bs{\vartheta}_{f})\bs{\Psi}(L,\bs{\vartheta}_{0,1})\bs{B}(\bs{\vartheta}_{0,2})\bs{\varepsilon}_{t}=\bs{\delta}(L,\bs{\vartheta}_{f},\bs{\vartheta
		}_{0})\bs{\varepsilon}_{t}.
	\end{equation}
	Notice that if the structural model in equation (\ref{eq:Eq1}) were fundamental, then $\bs{\delta}(L,\bs{\vartheta}_{f},\bs{\vartheta}_{0})=\bs{B}(\bs{\vartheta}_{0,2})$, which means that RF errors are a simply rotation of contemporaneous, unobserved structural shocks. Nonetheless, when only unit roots are discarded from AR or MA polynomials and the structural model is possibly non-fundamental, then $\bs{\delta}(L,\bs{\vartheta}_{f},\bs{\vartheta}_{0})$ is a non-causal filter. This implies that RF errors are not a static rotation of structural shocks, but a linear combination of leads and lags of the structural errors.
	
	For a simpler exposition, we write $\bs{\delta}(L,\bs{\vartheta}_{f},\bs{\vartheta}_{0})=\bs{\delta}(L,\bs{\vartheta}_{f})$. The no serial correlation of RF errors implies that
	\begin{equation}
		\label{eq:All-Pass}
		\bs{\delta}(e^{i\lambda},\bs{\vartheta}_{f})\bs{\delta}^{*}(e^{i\lambda},\bs{\vartheta}_{f}) = \bs{\Omega} 
	\end{equation}
	where $\bs{\Omega}$ is positive definite, symmetric, time-invariant matrix; $\bs{\delta}^{*}(e^{i\lambda},\bs{\vartheta}_{f})$ denotes the transposed, conjugate of matrix polynomial $\bs{\delta}(e^{i\lambda},\bs{\vartheta}_{f})$. \citet{velasco2022identification,baggio2018parametrization} and the references therein denominate filters satisfying condition (\ref{eq:All-Pass}) as \textit{all-pass} filters.\footnote{To be completely precise, an \emph{all-pass} filter requires that $\bs{\Omega}=\bs{I}_{d}$. However, this is not a problem in our analysis because $\bs{\Omega}$ is a constant matrix.}
	
\section{Higher-order Spectrum and non-Gaussian Dimension}
\label{sec:HOSA_CumRandVec}
	
	In order to determine the non-Gaussian dimension in the vectors of structural shocks $\bs{\varepsilon}_{t}$ and RF errors vector $\bs{u}_{t}$ it is necessary to characterize non-Gaussian behavior. Like \citet{jarque1987test,lobato2004simple,guay2021identification}, we typify non-Gaussianity as deviations of the higher order information respect to the their values under Gaussianity. And for assessing higher order information we exploit third and fourth order cumulants.

\subsection{Cumulants of Random Vector}
\label{sec:CumRandVector}
The cumulant generating function of vector $\bs{\varepsilon}_{t}$ is $\bs{\kappa}_{\bs{\varepsilon}}(\bs{\tau})\defeq\log\left(\phi_{\bs{\varepsilon}}(\bs{\tau})\right)$, where $\phi_{\bs{\varepsilon}}(\bs{\tau})\defeq\mathbb{E}(\exp{(i\bs{\tau}^{\prime}\bs{\varepsilon})})$ represents the characteristic function  of $\bs{\varepsilon}_{t}$ and $\bs{\tau}=(\tau_{1},\dots,\tau_{d})^{\prime}$ a $d$-dimensional real-valued vector. A cumulant of order $k$ (or the cumulant of a $k$-tuple $\{\varepsilon_{j_{m},t}\}_{m=1}^{k}$ ) is defined as
\begin{equation}
	\label{eq:Cumulant_sOrder}
	\Cum\left( \varepsilon_{j_{1},t},\dots,\varepsilon_{j_{k},t} \right) = \left. \frac{\partial^{k}}{(\partial{\tau_{1}})^{k_{1}}\dots(\partial{\tau_{d}})^{k_{d}}}{\bs{\kappa}_{\bs{\varepsilon}}(\bs{\tau})} \right|_{\bs{\tau}=\bs{0}},
\end{equation}
where $\sum_{j=1}^{d}{k_{j}}=k\geq1$ and $k_{j}\geq{0}$. $j_{m}\in\{1,\dots,d\}$ for all $m=1,\dots,k$. When moments (up to order $k$) exist, \citet{jammalamadaka2006higher} shows that the expression in equation (\ref{eq:Cumulant_sOrder}) is equal to
\begin{equation}
	\label{eq:Cumulant_sOrder_Moments}
	\Cum\left( \varepsilon_{j_{1},t},\dots,\varepsilon_{j_{k},t} \right) = \sum_{\mathfrak{p}}{(\abs{\mathfrak{p}} - 1)!(-1)^{\abs{\mathfrak{p}} - 1} \prod_{P\in\mathfrak{p}}{\E\left( \prod_{j_{m}\in{P}}{\varepsilon_{j_{m},t}} \right)} },
\end{equation}
where $\mathfrak{p}$ is a partition of set $\{j_{1},\dots,j_{k}\}$, $\abs{\mathfrak{p}}$ represents the number of parts in partition $\mathfrak{p}$. $P$ is an element of $\mathfrak{p}$. Although the expression in (\ref{eq:Cumulant_sOrder_Moments}) seems difficult to handle it, let see some simple examples. When $k=2$ and we choose the duplet $(\varepsilon_{1,t},\varepsilon_{2,t})$, then there exists only two partitions of set $\{1,2\}$, which are $\mathfrak{p}_{1}=\{1,2\}$ and $\mathfrak{p}_{2}=\{\{1\},\{2\}\}$. Thus $\Cum(\varepsilon_{1,t},\varepsilon_{2,t})=\E[\varepsilon_{1,t}\varepsilon_{2,t}]-\E[\varepsilon_{1,t}]\E[\varepsilon_{2,t}]=\Cov(\varepsilon_{1,t},\varepsilon_{2,t})$. 

When $k=3$ and the chosen triplet is $(\varepsilon_{1,t},\varepsilon_{2,t},\varepsilon_{3,t})$, there are five possible partitions $\mathfrak{p}_{1}=\{1,2,3\}$, $\mathfrak{p}_{2}=\{\{1\},\{2,3\}\}$, $\mathfrak{p}_{3}=\{\{2\},\{1,3\}\}$, $\mathfrak{p}_{4}=\{\{3\},\{1,2\}\}$ and $\mathfrak{p}_{5}=\{\{1\},\{2\},\{3\}\}$. Thus, $\Cum(\varepsilon_{1,t},\varepsilon_{2,t},\varepsilon_{3,t})=\E[\varepsilon_{1,t}\varepsilon_{2,t}\varepsilon_{3,t}]-\E[\varepsilon_{1,t}]\E[\varepsilon_{2,t}\varepsilon_{3,t}]-\E[\varepsilon_{2,t}]\E[\varepsilon_{1,t}\varepsilon_{3,t}]-\E[\varepsilon_{3,t}]\E[\varepsilon_{1,t}\varepsilon_{2,t}]+2\E[\varepsilon_{1,t}]\E[\varepsilon_{2,t}]\E[\varepsilon_{3,t}]=\E[\varepsilon_{1,t}\varepsilon_{2,t}\varepsilon_{3,t}]$. 

When the elements in the $k$-tuple satisfies $j_{1}=j_{2}=\cdots=j_{k}=m$, then we call $\Cum\left( \varepsilon_{m,t},\dots,\varepsilon_{m,t} \right)$ the $m$-th marginal cumulant of order $k$, and it is represented by $\kappa_{k,m}^{\varepsilon}$. Collecting all the cumulants of order $k$ in a single-column vector, we obtain
\begin{equation}
	\label{eq:CumulantVector}
	\bs{\kappa}_{k}(\bs{\varepsilon}) = \left[ \Cum(\varepsilon_{j_{1},t},\dots,\varepsilon_{j_{k},t}) \right]_{(j_{1},\dots,j_{k})\in\bs{\upsigma}_{k}(\{1,\dots,d\})}
\end{equation}
where $\bs{\upsigma}_{k}(\{1,\dots,d\})$ is the set of all permutations of length $k$ formed with numbers in the set $\{1,\dots,d\}$. For instance, if $d=2$ and $k=2$, $\bs{\upsigma}_{2}(\{1,\dots,2\})=\left\{ \{1,1\}; \{1,2\}; \{2,1\}; \{2,2\} \right\}$. From expression in (\ref{eq:CumulantVector}), it clear that $\bs{\kappa}_{k}(\bs{\varepsilon})$ is $d^{k}$-dimensional real-valued vector. The number of non-repeated elements in $\bs{\kappa}_{k}(\bs{\varepsilon})$ is $\binom{d+k-1}{k}=\frac{(d+k-1)!}{k!(d-1)!}$.

Let $\mb{v}_{k}^{\bs{\varepsilon}}$ be a $d^{k-1}\times{d}$ matrix such that $\bs{\kappa}_{k}(\bs{\varepsilon})=\vect(\mb{v}_{k}^{\bs{\varepsilon}})$, i.e, $\mb{v}_{k}^{\bs{\varepsilon}}$ is a matrix version of vector $\bs{\kappa}_{k}(\bs{\varepsilon})$. By assumptions (\ref{as:Assumption1}.(ii)-(iv)), all non-marginal cumulants of $\bs{\varepsilon}_{t}$ are zero, then $\mb{v}_{k}^{\bs{\varepsilon}}$ has the following structure
\begin{align}
	\label{eq:CumulantMatrixVersion}
	\mb{v}_{k}^{\bs{\varepsilon}} &= \begin{bmatrix}
		\kappa_{k,1}\bs{e}_{1}^{\otimes{k-1}} & \kappa_{k,2}\bs{e}_{2}^{\otimes{k-1}} & \dots & \kappa_{k,d}\bs{e}_{d}^{\otimes{k-1}}
	\end{bmatrix},
\end{align}
where $\bs{e}_{j}$ is the $j$-th canonical vector in $\mbb{R}^{d}$.

\subsection{Number of non-Gaussian shocks}
Since we are characterizing non-Gaussian behavior by its deviation from Gaussianity using third and fourth order information, then the number of non-Gaussian distributed components in the vector of structural shocks $\bs{\varepsilon}$ is equal to number of non-zero third or fourth order marginal cumulants $\kappa_{k,m}^{\varepsilon}$. And, from expression (\ref{eq:CumulantMatrixVersion}), the number of non-zero third order cumulants is equal to the rank of matrix $\mb{v}_{3}^{\bs{\varepsilon}}$; while the number of non-zero fourth order cumulants is $\rank(\mb{v}_{4}^{\bs{\varepsilon}})$. Besides, the rank of the block matrix $\mb{v}_{34}^{\bs{\varepsilon}}=\begin{bmatrix}
	\mb{v}_{3}^{\bs{\varepsilon}} \\ \mb{v}_{4}^{\bs{\varepsilon}}
\end{bmatrix}$ gives the joint number of non-zero third and fourth order cumulants. Therefore, it can be said that $\rank(\mb{v}_{3}^{\bs{\varepsilon}})$ reveals the amount of asymmetric shocks, $\rank(\mb{v}_{4}^{\bs{\varepsilon}})$ provides the number of non-mesokurtic shocks.

This result is behind the approach followed by \citet{maxand2020identification} work. Given that her work restricts the focus to a causal (fundamental) SVAR model, RF errors are a simple rotation of contemporaneous value of structural shocks. And, even though the whole vector of structural errors is not identified, she demonstrates that the non-Gaussian block it is. Therefore, she employs \citet{nordhausen2017asymptotic} strategy for estimating the rank or the non-Gaussian dimension.

\citet{guay2021identification} works with a similar context as \citet{maxand2020identification}, i.e., assuming a causal (fundamental) SVAR as the generating process for observables. But, unlike Maxand's work, he does not exploit $\mb{v}_{k}^{\bs{\varepsilon}}$ directly. As we explained above, the RF errors, under fundamentalness, are a linear rotation of structural shocks, that is $\bs{u}_{t}=\bs{B}(\bs{\vartheta}_{0,2})\bs{\varepsilon}_{t}$ where $\bs{\vartheta}_{0,2}$ governs the contemporaneous or static part of the fundamental SVAR model and is not identified. According to \citet{jammalamadaka2006higher}, the vector of $k$-th order cumulant of the vector of RF errors is $\bs{\kappa}_{k}^{\bs{u}} = \bs{B}^{\otimes{k}}\bs{\kappa}_{k}({\bs{\varepsilon}})$ where $\bs{B}^{\otimes{k}}$ denotes the $k$-th Kronecker power.\footnote{Notice that we are employing $\bs{B}\defeq\bs{B}(\bs{\vartheta}_{0,2})$.} Then, similarly to $\mb{v}_{k}^{\bs{\varepsilon}}$ we define
\begin{align*}
	\mb{v}_{k}^{\bs{u}} \defeq \vect^{-1}(\bs{\kappa}_{k}^{\bs{u}}) &= \bs{B}^{\otimes{k-1}}\mb{v}_{k}^{\bs{\varepsilon}}\bs{B}^{\prime}
\end{align*}
Since $\bs{B}$ is a square, full rank matrix, then $\rank\left( \mb{v}_{k}^{\bs{u}} \right)=\rank\left( \mb{v}_{k}^{\bs{\varepsilon}} \right)$. Therefore, it can be concluded that under fundamentalness, the non-Gaussian dimension in the structural model can be identified by calculating the rank of a matrix constructed with third or fourth cumulants of RF errors.

Nonetheless, under a more general structural model like the one considered in this paper, \citet{maxand2020identification} and \citet{guay2021identification} approaches are invalid. First, a possibly non-fundamental structural VARMA is only identified if all shocks are non-Gaussian, which we ignore. Hence, it would be no reliable any approach that employs estimated structural shocks. Second, as it showed in the expression (\ref{eq:ReducedFormErrors}), RF errors from a fundamental VARMA model are a filtered version of structural shocks. In this case, $\bs{\kappa}_{k}^{\bs{u}}$ and $\mb{v}_{k}^{\bs{u}}$ are
\begin{align}
	\bs{\kappa}_{k}^{\bs{u}} &= \sum_{j=-\infty}^{\infty}{\bs{\delta}_{j}^{\otimes{k}}\bs{\kappa}_{k}^{\bs{\varepsilon}}} \nonumber \\
	\label{eq:Cumulants_RFerrors_Matrix}
	\mb{v}_{k}^{\bs{u}} &= \sum_{j=-\infty}^{\infty}{\bs{\delta}_{j}^{\otimes{k-1}}\mb{v}_{k}^{\bs{\varepsilon}}\bs{\delta}_{j}^{\prime} }
\end{align}
From this latter expression, it is visible that $\rank\left( \mb{v}_{k}^{\bs{u}} \right)\neq \rank\left( \mb{v}_{k}^{\bs{\varepsilon}} \right)$ in general, because even though the rank of $\bs{\delta}_{j}^{\otimes{k-1}}\mb{v}_{k}^{\bs{\varepsilon}}\bs{\delta}_{j}^{\prime}$ may coincide with $\rank\left(\mb{v}_{k}^{\bs{\varepsilon}}\right)$, this term may be positive or negative definite. Thus, the rank of the summation could be higher or lower than $\rank\left(\mb{v}_{k}^{\bs{\varepsilon}}\right)$.\footnote{Notice a simple example of having two matrices $A=\begin{bmatrix}
		1 & 0 \\ 0 & 1
\end{bmatrix}$ and $B=\begin{bmatrix}
-1 & 0 \\ 0 & 1
\end{bmatrix}$. Both are full rank matrices, but the sum $C=A+B=\begin{bmatrix}
0 & 0 \\ 0 & 2
\end{bmatrix}$ is rank-deficient.} Consequently, the rank of $\mb{v}_{k}^{\bs{u}}$ does not identify anymore our parameter of interest, $\rank(\mb{v}_{k}^{\bs{\varepsilon}})$.

To overcome this issue and instead of using only third or fourth information of contemporaneous RF errors, we exploit the third and fourth information at all leads and lags of RF errors. Hence, we employ the $k$-th order cumulant spectral density of the RF errors. According to \citet{brillinger2001time}, the cumulant spectrum of order $k$ is
\begin{align}
	\label{eq:EqHOSA_Spectrum}
	g_{k}^{u}(\bs{\lambda}) &= (2\pi)^{k-1}\left(\bs{\delta}\left(e^{i\sum_{m=1}^{k-1}{\lambda}_{m}};\bs{\vartheta}_{f}\right)\otimes \bigotimes_{j=1}^{k-1}\left\{ \bs{\delta}\left(e^{-i\lambda_{k-j}};\bs{\vartheta}_{f}\right) \right\}\right)\bs{\kappa}_{k}\left(\bs{\varepsilon}\right),
\end{align}
where $\bigotimes_{j=1}^{n}{\bs{A}_{j}}\defeq\bs{A}_{1}\otimes\bs{A}_{2}\otimes\cdots\otimes{\bs{A}_{n}}$. 

Similarly to $\mb{v}_{k}^{\bs{u}}$, we define the $d^{k-1}\times{d}$ matrix $G_{k}^{u}(\bs{\lambda})$, such that  $g_{k}^{u}(\bs{\lambda})=\vect\left( G_{k}^{u}(\bs{\lambda}) \right)$. Thus
\vspace*{-0.25cm}
\begin{align}
	\label{eq:EqHOSA_Spectrum_Matrix}
	G_{k}^{u}(\bs{\lambda}) &= \left( \bigotimes_{j=1}^{k-1}\left\{ \bs{\delta}\left(e^{-i\lambda_{k-j}};\bs{\vartheta}_{f}\right) \right\} \right)\mb{v}_{k}^{\bs{\varepsilon}}\left(\bs{\delta}^{\prime}\left(e^{i\sum_{m=1}^{k-1}{{\lambda}_{m}}}  ;\bs{\vartheta}_{f}\right)\right).
\end{align}

Based on $G_{k}^{u}(\bs{\lambda})$, let define the $d\times{d}$ matrix $G_{k}^{u,2}(\lambda_{k})\defeq\left[G_{k}^{u}(\bs{\lambda})\right]^{*}G_{k}^{u}(\bs{\lambda})$ where $\bs{A}^{*}$ denotes the conjugate transpose of the complex-valued matrix $A$. Since $\bs{\delta}(z,\bs{\vartheta}_{f})$ is an all-pass filter and using $(A\otimes{B})(C\otimes{D})=\left(AC\otimes{BD}\right)$, provided that $AC$ and $BD$ are conformable for multiplication, we obtain
\begin{align}
	\label{eq:GMat_Sq}
	\hspace{-0.5cm}
	G_{k}^{u,2}(\lambda_{k}) & = \bs{\delta}\left(e^{i\lambda_{k} };\bs{\vartheta}_{f}\right)\left(\mb{v}_{k}^{\bs{\varepsilon}\prime}\bs{\Omega}^{\otimes{(k-1)}}\mb{v}_{k}^{\bs{\varepsilon}}\right)\bs{\delta}^{\prime}\left(e^{-i\lambda_{k}};\bs{\vartheta}_{f}\right),
\end{align}
with $\lambda_{k}=-\sum_{m=1}^{k-1}{{\lambda}_{m}}$. Given that $G_{k}^{u,2}(\lambda_{k})$ only depends on the scalar frequency $\lambda_{k}\in[-\pi,\pi]$, thus, from now on, we can drop the sub-index $k$. 

Since $\bs{\delta}(z,\bs{\vartheta}_{f})$ is full rank for any $z\in\mathbb{C}$ such that $\abs{z}=1$, then it is clear that $\rank\left( G_{k}^{u,2}(\lambda) \right) = \rank\left( \mb{v}_{k}^{\bs{\varepsilon}\prime}\bs{\Omega}^{\otimes{k-1}}\mb{v}_{k}^{\bs{\varepsilon}} \right)=\rank\left(\mb{v}_{k}^{\bs{\varepsilon}\prime}\mb{v}_{k}^{\bs{\varepsilon}}\right)$. In consequence, the non-Gaussian dimension in $\bs{\varepsilon}_{t}$ can be identified by  $\rank\left( G_{k}^{u,2}(\lambda) \right)$. The following proposition summarizes our main finding.

\begin{proposition}\mbox{}\newline
	\label{prop:Proposition1}
	Let the structural model be described by equations (\ref{eq:Eq1}) and (\ref{eq:Eq2}), with structural shocks satisfying Assumption (\ref{as:Assumption1}). Consider the arrays $G_{3}^{u,2}(\lambda)$, $G_{4}^{u,2}(\lambda)$ and $G_{34}^{u,2}(\lambda) = \begin{bmatrix} G_{3}^{u,2}(\lambda) \\ G_{4}^{u,2}(\lambda)  \end{bmatrix}$, therefore
	\vspace*{-0.5cm}
	\begin{align*}
		\rank\left( G_{3}^{u,2}(\lambda) \right) &= \rank\left(\mb{v}_{3}^{\bs{\varepsilon}\prime}\mb{v}_{3}^{\bs{\varepsilon}}\right) = d_{3}, \\
		\rank\left( G_{4}^{u,2}(\lambda) \right) &= \rank\left(\mb{v}_{4}^{\bs{\varepsilon}\prime}\mb{v}_{4}^{\bs{\varepsilon}}\right) = d_{4},\\
		\rank\left(G_{34}^{u,2}(\lambda)\right) &= d_{34},\quad \forall\;\; \lambda\in[-\pi,\pi],
	\end{align*}
	where $d_{3}$ and $d_{4}$ are the number of skewed and non-mesokurtic structural shocks, respectively, and $d_{34}$ is the number of asymmetric or non-mesokurtic shocks.
\end{proposition}
Proposition (\ref{prop:Proposition1}) can be interpreted as follows: the non-Gaussian dimension in the vector of structural shocks is equal to the rank of a matrix constructed from higher order cumulant spectrum of RF errors at a given frequency. In particular, if only the third-order cumulant spectrum is employed, we obtain the non-Gaussian dimension delivered by asymmetric non-Gaussian shocks. If only the fourth-order spectrum is utilized, we capture the non-Gaussian dimension spanned by non-mesokurtic shocks.

Besides, notice that in case the condition in (\ref{eq:Eq2}) holds for all $z\in\mbb{T}_{+}=\{x\in\mbb{C}\;|\;\abs{x}\leq{1}\}$, higher order cumulant spectrums are constant. Hence, array $G^{u,2}_{k}(\lambda)$ is constant, i.e., the same for any frequency. In particular, $G^{u,2}_{k}(\lambda)=\bs{B}(\bs{\vartheta}_{0,2})\left(\mb{v}_{k}^{\bs{\varepsilon}\prime}\bs{\Omega}^{\otimes{k-1}}\mb{v}_{k}^{\bs{\varepsilon}}\right)\bs{B}^{\prime}(\bs{\vartheta}_{0,2})=\mb{v}_{k}^{\bs{u}\prime}\mb{v}_{k}^{\bs{u}}$ for any $\lambda\in[-\pi,\pi]$. This setup is the case analyzed in \cite{guay2021identification}. Under this situation, proposition (\ref{prop:Proposition1}) holds as well. In other words, if a fundamental structural VARMA model generates $\bs{y}_{t}$, a rectangular array constructed from cumulant spectrums of order $3$ or $4$ based on RF errors identifies the non-Gaussian dimension.
			
	\section{Estimating the number of non-Gaussian elements}
	\label{sec:EstimatingNumber}
From the discussion above, to determine the non-Gaussian dimension in the structural shocks vector, we need to find the rank of a rectangular array constructed from the cumulant spectrum of order $k=3,4$. Thus, our empirical problem becomes the estimation of the rank of a matrix. In the literature, this problem has been dealt with in two approaches: sequential testing or estimation by information criteria (see \citet{camba2009statistical} for a detailed review). We opt for the former approach because it may be problematic to construct a pseudo-likelihood function for a non-causal filter. Additionally, the parametric estimation of our matrix of interest is not feasible because the possibly non-fundamental filter $\bs{\delta}(z,\bs{\vartheta}_{f},\bs{\vartheta}_{0,1})=\bs{\Psi}^{-1}(z,\bs{\vartheta}_{f})\bs{\Psi}(z,\bs{\vartheta}_{0,1})$ is not identified unless we impose fundamentalness. Consequently, we decide to estimate $g_{k}(\bs{\lambda})$ non-parametrically.

\subsection{Estimation of non-Gaussian Dimension}
\label{sec:TestRank}
Dealing with $G_{k}^{u,2}(\lambda)$ may be problematic since it implies working with complex-valued terms. For simplifying the analysis, we use $\text{Re}(G_{k}^{2}(\lambda))$ instead. At frequency zero, there is no problem since $G_{k}^{u,2}(0)=\text{Re}(G_{k}^{u,2}(0))$. At other frequencies, $\text{Re}(G_{k}^{u,2}(\lambda))$ is the sum of two positive semidefinite quadratic forms; thus, we cannot lose rank, and we gain rank only if both quadratic forms has linearly independent columns. 

We now describe our estimation of the non-Gaussian dimension in structural shocks through a sequential hypothesis test. At step $s=1,2,\dots,d$ in the sequential procedure, the null and alternative hypotheses are
\begin{align*}
	H_{0,s}: & \rank\left( \text{Re}(G_{k}^{u,2}(\lambda)) \right) = r_{s} \\
	H_{1,s}: & \rank\left( \text{Re}(G_{k}^{u,2}(\lambda)) \right) \geq  r_{s}+1,
\end{align*}
where $r_{s}$ denotes the rank under the null hypothesis at step $s$. 

We start the sequential procedure (at step $s=1$) by imposing $r_{1}=0$, i.e., joint Gaussianity of the structural shocks. If this is rejected, we continue with the next step, $s=2$, and the null hypothesis is updated to $r_{2}=1$, i.e., only one asymmetric or non-mesokurtic structural shock. We continue this way until we cannot reject a null hypothesis or reach the final step $s=d$, where $r_{d}=d-1$. If this is not rejected, the non-Gaussian dimension in the structural shocks vector is $d-1$; on the contrary, the non-Gaussian dimension is $d$.

The literature on rank estimation via hypothesis testing is extensive (see \citet{al2017unifying} for a complete survey). We center our approach on \citet{kleibergen2006generalized} (hereafter KP) proposal. They employ the singular value decomposition (hereafter SVD) of a matrix of interest, $\bs{\Pi}$. The SVD of matrix $\bs{\Pi}_{m\times{n}}$ consists on finding squared orthonormal matrices $\bs{R}_{1}$ and $\bs{R}_{2}$ of dimension $m$ and $n$, respectively; and a quasi-diagonal rectangular matrix $\bs{\mathcal{L}}_{m\times{n}}$ such that\footnote{See \citet{golub2012matrix} for more details on SVD of a matrix.}
\begin{equation}
	\label{eq:SVD}
	\bs{R}_{1}^{\prime}\bs{\Pi}\bs{R}_{2}=\bs{\mathcal{L}},
\end{equation}
where $\bs{\mathcal{L}}$ is a rectangular array with a squared block  $\bs{\mathcal{L}}_{\longbar{m}}=\text{diag}(l_{1},\dots,l_{\longbar{m}})$ with $\longbar{m}=\min\{m,n\}$ and $l_{1}\geq{l_{2}}\geq\dots\geq{l_{\longbar{m}}}\geq{0}$ are the singular values of $\bs{\Pi}$. If $\bs{\Pi}$ is squared, then $\bs{\mathcal{L}}=\bs{\mathcal{L}}_{\longbar{m}}$. Assuming that $m>n$, $\bs{\mathcal{L}}=\begin{bmatrix}
	\bs{\mathcal{L}}_{\longbar{m}} & \bs{0}_{n\times{(m-n)}}
\end{bmatrix}^{\prime}$.

From decomposition in (\ref{eq:SVD}) and selecting an integer $r\in\{0,\dots,\longbar{m}-1\}$, \citet{kleibergen2006generalized} obtain that
\begin{equation}
	\label{eq:SVD_KP}
	\bs{\Pi} = \bs{C}_{r}\bs{D}_{r} + \bs{C}_{r,\perp}\bs{\mathcal{L}}_{r}\bs{D}_{r,\perp}
\end{equation}
where $\bs{C}_{r}\bs{D}_{r}=\underbrace{\begin{bmatrix}\bs{R}_{1,11} \\ \bs{R}_{1,21}\end{bmatrix}}_{m\times{r}}\underbrace{\bs{\mathcal{L}}_{1}}_{r\times{r}}\underbrace{\begin{bmatrix}\bs{R}_{2,11}^{\prime} & \bs{R}_{2,21}^{\prime}\end{bmatrix}}_{r\times{n}}$; $\bs{C}_{r,\perp}\bs{\mathcal{L}}_{r}\bs{D}_{r,\perp}=\underbrace{\begin{bmatrix}\bs{R}_{1,12} \\ \bs{R}_{1,22}\end{bmatrix}}_{m\times{(m-r)}}\underbrace{\bs{\mathcal{L}}_{2}}_{(m-r)\times(n-r)}\underbrace{\begin{bmatrix}\bs{R}_{2,12}^{\prime} & \bs{R}_{2,22}^{\prime}\end{bmatrix}}_{(n-r)\times{n}}$. $\bs{C}_{r,\perp}$ and $\bs{D}_{r,\perp}$ are the orthogonal complements of $\bs{C}_{r}$ and $\bs{D}_{r}$, respectively.  Thus, \citet{kleibergen2006generalized} approach consists in decomposing the matrix $\bs{\Pi}$ into two linear independent parts: one that is made of the multiplication of two matrices that are full column rank, $\bs{C}_{r}\bs{D}_{r}$, and another with null rank $\bs{C}_{r,\perp}\bs{\mathcal{L}}_{r}\bs{D}_{r,\perp}$. Therefore, $\rank(\bs{\Pi})=\rank(\bs{C}_{r}\bs{D}_{r})=r$. This result is only achieved if we place all the non-zero singular values into $\bs{\mathcal{L}}_{1}$. Accordingly, the null hypothesis of $\rank(\bs{\Pi})=r$ is equivalent to $\bs{\mathcal{L}}_{r}=\bs{0}$. 

Since $\bs{\Pi}$ is not observable, the SVD is applied to its estimator $\bs{\widehat{\Pi}}$. Thus, we can write $\bs{\widehat{\Pi}}=\bs{\widehat{C}}_{r}\bs{\widehat{D}}_{r} + \bs{\widehat{C}}_{r,\perp}\bs{\widehat{\mathcal{L}}}_{r}\bs{\widehat{D}}_{r,\perp}$. Under the assumption of $\sqrt{T}(\vect{[\bs{\widehat{\Pi}}]}-\vect[\bs{\Pi}]) \xrightarrow{d} \mathcal{N}_{mn}(\bs{0},\bs{\Xi})$ with $\bs{\Xi}$ positive definite, then under the null hypothesis of $\rank(\bs{\Pi})=r$ the asymptotic distribution of $\bs{\widehat{\ell}}_{r}=\vect\left(\bs{\widehat{\mathcal{L}}}_{r}\right)$ is $\sqrt{T}\bs{\widehat{\ell}}_{r} \xrightarrow{d} \mathcal{N}_{(m-r)(n-r)}\left(\bs{0}, \widetilde{\bs{\Xi}}_{r}\right)$ where $\bs{\widetilde{\Xi}}_{r}=(\bs{D}_{r,\perp}\otimes{\bs{C}_{r,\perp}^{\prime}})\bs{\Xi}(\bs{D}_{r,\perp}\otimes{\bs{C}_{r,\perp}^{\prime}})$. The KP-statistic is
\begin{equation}
	\label{eq:KP_Stat}	
	KP_{r}^{(T)}=T\bs{\widehat{\ell}}_{r}^{\prime}\widehat{\bs{\widetilde{\Xi}}}_{r}^{-1}\bs{\widehat{\ell}}_{r} \xrightarrow{d} \chi^{2}_{(m-r)(n-r)},
\end{equation}	
where $\widehat{\bs{\widetilde{\Xi}}}_{r}$ is the consistent estimator of the asymptotic variance, ${\bs{\widetilde{\Xi}}}_{r}$.

In some dimensions, our problem departs from \citet{kleibergen2006generalized} context. First, our matrix of interest $\bs{\Pi}=\text{Re}({G}_{k}^{u,2}(\lambda))$ is symmetric, i.e. it contains repeated elements. Hence, the asymptotic variance of unrestricted estimator $\bs{\widehat{\Pi}}$, $\bs{\Xi}$, is only positive semi-definite. Second, as we detail in the section where it is discussed the estimation of higher order spectrum, the asymptotic distribution of the unrestricted estimator of our matrix of interest changes under the null hypothesis of joint Gaussianity, i.e., under the null of $\rank(\bs{\Pi})=0$. Finally, our convergence rates for the asymptotic distribution of the statistic are lower in comparison to the standard speed of convergence, $\sqrt{T}$.

\paragraph{Test Statistic at first step}\mbox{}\newline
For the first step in the sequential procedure, the null hypothesis is $\rank(\bs{\Pi})=0$, i.e., the vector of structural shocks is an uncorrelated Gaussian process. The KP statistic under this null hypothesis is
\begin{align}
	\hspace{-0.5cm}
	KP^{(T)}_{r_{1}} =a_{T}^{4}\widehat{\bs{\ell}}_{0}^{\prime}\widehat{\text{aVar}}\left( \left( \overline{\bs{D}}_{0,\perp}\otimes\overline{\bs{C}}^{\prime}_{0,\perp} \right)\bs{Q} \right)^{\dagger}\widehat{\bs{\ell}}_{0},
\end{align}
where $\bs{A}^{\dagger}$ denotes the Moore-Penrose inverse of matrix $\bs{A}$, $\bs{Q}$ is the vectorized form of a linear combination of an inverse Wishart distribution, which parameters depends on the asymptotic variance of spectrum estimates (see Section $4.3$ below), and $ \overline{\bs{D}}_{0,\perp}$ and $\overline{\bs{C}}^{\prime}_{0,\perp}$ represent the limiting values of $\widehat{\bs{D}}_{0,\perp}$ and $\widehat{\bs{C}}_{0,\perp}$. $a_{T}$ is the convergence rate of spectrum estimates.

\paragraph{Test Statistic under mixed cases}\mbox{}\newline
Now, when the null hypothesis is $\rank(\bs{\Pi})=r$ with $r>{0}$, i.e., the vector of structural shocks is an i.i.d process with mutually independent components and there are $r$ non-Gaussian distributed shocks. In this case, the asymptotic distribution of $\bs{\widehat{\Pi}}$ is Normal, and the KP statistic adopts a similar form as in \citet{kleibergen2006generalized},
\begin{equation}
	\label{eq:KP_Stat2}
	KP^{(T)}_{r_{s}}=a_{T}^{2}\bs{\widehat{\ell}}_{r}^{\prime}\left(\left[(\widehat{\bs{D}}_{r,\perp}\otimes{\widehat{\bs{C}}_{r,\perp}^{\prime}})\widehat{\bs{\Xi}}(\widehat{\bs{D}}_{r,\perp}\otimes{\widehat{\bs{C}}_{r,\perp}^{\prime}})^{\prime}\right]\right)^{\dagger}\bs{\widehat{\ell}}_{r}.
\end{equation}
The statistic in (\ref{eq:KP_Stat2}) is asymptotically distributed as a chi-square with degrees of freedom $\nu=\rank{\left[(\widehat{\bs{D}}_{r,\perp}\otimes{\widehat{\bs{C}}_{r,\perp}^{\prime}})\widehat{\bs{\Xi}}(\widehat{\bs{D}}_{r,\perp}\otimes{\widehat{\bs{C}}_{r,\perp}^{\prime}})^{\prime}\right]}$. Unfortunately, the exact value for $\nu$ is not easy to determine, because -although $\widehat{\bs{C}}_{r,\perp}$ and  $\widehat{\bs{D}}^{\prime}_{r,\perp}$ are full column rank matrices of dimension $d\times{(d-r)}$- the matrix $(\widehat{\bs{D}}_{r,\perp}\otimes{\widehat{\bs{C}}_{r,\perp}^{\prime}})$ is full row rank matrix of dimension $(d-r)^{2}\times{d^{2}}$ and the matrix $\widehat{\bs{\Xi}}$ is only positive semidefinite, i.e. it is not full rank. From this discussion, we can bound the degrees of freedom, $1\leq \nu\leq \min\{ \rank(\bs{\Xi}), (d-r_{s})^{2} \}$.

\subsection{Bootstrap Test}
Because of the difficulties characterizing the asymptotic distributions of the KP statistic for each step in the sequential procedure, we opt for a bootstrap strategy. Nonetheless, this path entails other challenges. The most problematic is to accomplish that the bootstrap sample appropriately reflects the null hypothesis of each step. Otherwise, it may severely compromise the size and power of the test (see \citet{hall1991two,portier2014bootstrap}).

It is a well-known result from the SVD of a matrix $\bs{\Pi}_{m\times{n}}$ with $m>n$ and $\rank(\bs{\Pi})=r$, that $\text{null}(\bs{\Pi})$ is spanned by the last $n-r$ columns of $\bs{R}_{2}$ and the range $\text{ran}(\bs{\Pi})$ is spanned by the first $r$ columns of $\bs{R}_{1}$. Thus, when $\rank(\bs{\Pi})=r$, the first $r$ columns of $\bs{R}_{1}$ span the non-Gaussian dimension, while last $n-r$ columns of $\bs{R}_{2}$ span the Gaussian one. Thus, we follow \citet{nordhausen2017asymptotic} approach and use the estimates of $\bs{R}_{2}$ for a projection of residuals into Gaussian and non-Gaussian dimensions.

Given the sample $\{\bs{y}_{t}\}_{t=1}^{T}$, we estimate a RF-VARMA model and compute its RF residuals $\{\hat{\bs{u}}_{t}\}_{t=1}^{T}$. Based on these estimates, we construct our matrix of interest $\widehat{\bs{\Pi}}=\text{Re}\left(\widehat{G}_{k}^{\Hat{u},2,(T)}(\lambda)\right)$ and the matrices $\widehat{\bs{\mathcal{L}}}$, $\widehat{\bs{R}}_{1}$ and $\widehat{\bs{R}}_{2}$. Under $H_{0,r}:\,\rank(\bs{\Pi})=r$ and according to \citet{nordhausen2017asymptotic}, we construct $\widehat{\bs{R}}_{2,r}\defeq\left[ \hat{\mb{r}}_{2,r+1}\cdots{\hat{\mb{r}}_{2,d} } \right]$ and the projection matrix $\mb{M}_{r}=\bs{I}-\hat{\Sigma}_{\hat{u}}^{1/2}\widehat{\bs{R}}_{2,r}\widehat{\bs{R}}_{2,r}^{\prime}\hat{\Sigma}_{\hat{u}}^{-1/2}$, where $\hat{\Sigma}_{\hat{u}}$ is variance of RF residuals vector. $\mb{M}_{r}$ is the projection matrix into the orthogonal space to the one spaned by $\widehat{\bs{R}}_{2,r}$, i.e., $\mb{M}_{r}$ is a projection into non-Gaussian space. Based on these inputs, at any step $s$ in the sequential testing procedure, each bootstrap sample is created following Algorithm 1.
\begin{algorithm}[!tbh]
	\caption{Bootstrap Sample under $H_{0,s}:\;\rank(\bs{\Pi})=r_{s}$}\label{alg:Algorithm_1}
	\begin{enumerate}
		\item Obtain an unrestricted bootstrap sample of RF residuals $\{\tilde{\bs{u}}_{t}\}_{t=1}^{T}$.
		\item Draw a sequence of size $T$ of random vectors, $\{\bs{\eta}_{t}\}_{t=1}^{T}$ , from multivariate standard Normal distribution,  $\mathcal{N}(\bs{0},\bs{I}_{d-r_{s}})$.
		\item Bootstrap sample of RF residuals consistent with $H_{0,s}$ is given by
		$${\bs{u}}^{\star}_{t} = \mb{M}_{r_{s}}\tilde{\bs{u}}_{t} + \hat{\Sigma}_{\hat{u}}^{1/2}\widehat{\bs{R}}_{2,r_{s}}\bs{\eta}_{t}.$$
		\item The restricted bootstrap sample of $\{\bs{y}^{\star}_{t}\}_{t=1}^{T}$ is built parametrically, i.e. using ${\bs{u}}^{\star}_{t}$ and estimated parameters from RF-VARMA.
	\end{enumerate}
\end{algorithm}

It is worth mentioning that the unrestricted bootstrap sample of RF residuals is not performed by the usual \textit{independent bootstrap} procedure stated by \citet{efron1979bootstrap}, but by the proposed method in \citet{politis1994stationary}, called the \textit{stationary bootstrap}. According to the authors, this procedure is suitable for stationary weakly dependent time series because, unlike other proposals such as \citet{kunsch1989jackknife,politis1992general}, it exhibits the desirable property that the resampled time series obtained are stationary conditional on the original data. This type of bootstrap sampling is necessary because RF errors are only serially uncorrelated, but they are not independent unless the fundamentalness of data is imposed or Gaussian is assumed. Besides, for obtaining the restricted bootstrap sample for observable data, we employ a parametric approach; otherwise is not possible to impose the null hypothesis in the data.

Once the restricted bootstrap sample has been obtained and we set a nominal size $\alpha\in(0,1)$ for all the steps in the sequence, the test proceeds as follows:
\begin{algorithm}[!tbh]
	\caption{Bootstrap Test}
	\begin{enumerate}
		\item[0.]  Set the initial step $s=1$.
		\item Given the step $s$, set the null hypothesis as $H_{0,s}:\, \rank(\bs{\Pi})=r_{s}=s-1$ and compute the KP statistic with original sample, $KP_{r_{s}}^{(T)}$.
		\item Use Algorithm \ref{alg:Algorithm_1} for obtaining $B$ bootstrap samples. Compute statistic $KP_{r_{s},b}^{(T),\star}$ for each bootstrap sample $b=1,\dots,B$.
		\item Compute the bootstrap p-value for $H_{0,r_{s}}$ as $\widehat{pv}_{r_{s}}=\frac{1+\sum_{b=1}^{B}{\mb{1}\left( KP^{(T)}_{r_{s}}\leq KP^{(T),\star}_{r_{s},b} \right)}}{1+B}$ and we decide in this fashion:
		\begin{enumerate}
			\item[3.1.] If $\widehat{pv}_{r_{s}}\geq \alpha$, then $H_{0,s}$ is not rejected, the estimated rank is $\hat{r}=r_{s}$ and the procedure ends;
			\item[3.2.] else if $s<d-1$, update $s=s+1$ and repeat since stage 1;
			\item[3.3.] else, the procedure ends and the rank is $\hat{r}=d$.
		\end{enumerate}
	\end{enumerate}
\end{algorithm}

\subsection{Estimation of Cumulant Spectrum of order $k$}
This part briefly discusses some details of estimating higher order cumulant spectrum. The discussion follows closely \citet{brillinger1967asymptotic}. The sample periodogram for a $k$-tuple of RF errors is
\begin{equation}
	\label{eq:Periodogram_k}
	I_{\bs{c},k}^{(T)}(\bs{\lambda}) = \frac{1}{(2\pi)^{k-1}T}\prod_{j=1}^{k}{z_{c_{j}}^{(T)}(\lambda_{t_{j}})},
\end{equation}
where $\bs{c}=\left(c_{1},\dots,c_{k}\right)\in\upsigma_{k}\left(\{1,\dots,d\}\right)$. $\lambda_{t_{j}}=2\pi \frac{t_{j}}{T}$ for $t_{j}=1,\dots,T-1$, $j=1,\dots,k-1$ and $\sum_{j=1}^{k}{\lambda_{t_{j}}}=0[\text{mod}(2\pi)]$. Besides, $z_{c_{j}}^{(T)}(\lambda_{t_{j}})=\sum_{t=0}^{T-1}{u_{c_{j},t}e^{-i\lambda_{t_{j}}{t}}}$ is the discrete Fourier transform (DFT) of $c_{j}$-th reduced-form error, $u_{c_{j},t}$.

A consistent estimator of the cumulant spectrum of order $k$ for a $\bs{c}$-tuple of RF errors is
\begin{equation}
	\label{eq:SpectrumK}
	\hat{g}_{\bs{c},k}^{\Hat{u},(T)}(\bs{\lambda}) = \left(\frac{2\pi}{T}\right)^{k-1} \sum_{t_{1},\dots,t_{k}=0}^{T-1}{W_{T}\left( \lambda_{1}-\lambda_{t_{1}},\dots,\lambda_{k}-\lambda_{t_{k}} \right)I_{\bs{c},k}^{(T)}\left( \bs{\lambda} \right) }
\end{equation}
with $W_{T}(\bs{a})=H_{T}^{-(k-1)}\sum_{\bs{j}=(j_{1},\dots,j_{k}):\sum_{m=1}^{k}{(j_{m}+a_{m})}=0}{ W\left( \frac{1}{H_{T}}(\bs{a}+2\pi\bs{j}) \right) }$ and $H_{T}$ satisfies $\lim\limits_{T\to{\infty}}{H_{T}}=0$ and $\lim\limits_{T\to\infty}{TH_{T}^{k-1}}=\infty$. The weighting function, $W(\bs{a})$, is symmetric around $\bs{0}$ and satisfies conditions stated in \citet[Assumption II]{brillinger1967asymptotic}.

The consistent estimator of $g_{k}^{u}(\bs{\lambda})$ is the collection of all $\hat{g}_{\bs{c},k}^{\Hat{u},(T)}(\bs{\lambda})$ in a vector, which we call $\hat{g}_{k}^{\hat{u},(T)}(\bs{\lambda})$. Thus, we can obtain the $d^{k-1}\times{d}$ matrix $\hat{G}_{k}^{\Hat{u},(T)}(\bs{\lambda})$ such that $\hat{g}_{k}^{\Hat{u},(T)}(\bs{\lambda}) = \vect\left( \hat{G}_{k}^{\Hat{u},(T)}(\bs{\lambda}) \right)$. Besides, $\hat{G}_{k}^{\Hat{u},2,(T)}(\lambda)= \left[\hat{G}_{k}^{\Hat{u},(T)}(\bs{\lambda})\right]^{*}\hat{G}_{k}^{\Hat{u},(T)}(\bs{\lambda})$. Finally, our matrix of interest is $\text{Re}\left( \hat{G}_{k}^{\Hat{u},2,(T)}(\lambda) \right)$.

\subsection{Asymptotics of spectral estimators}
\citet{brillinger1967asymptotic} show that
\begin{align}
	\hspace*{-1.0cm}
	\label{eq:AsympSpectEstimator}
	\sqrt{H_{T}^{k-1}T}\left( \begin{bmatrix}
		\text{Re}(\hat{g}_{k}^{\Hat{u},(T)}(\bs{\lambda})) \\
		\text{Im}(\hat{g}_{k}^{\Hat{u},(T)}(\bs{\lambda}))
	\end{bmatrix} -  \begin{bmatrix}
		\text{Re}({g}_{k}^{u}(\bs{\lambda})) \\
		\text{Im}({g}_{k}^{u}(\bs{\lambda}))
	\end{bmatrix} \right) &\xrightarrow{d} \mathcal{N}_{2d^{k}}\left(\bs{0}, \bs{\mathcal{V}}(\bs{\lambda})\right)
\end{align}
where $\mathcal{N}_{2d^{k}}$ represents a multivariate Gaussian random $2d^{k}$-dimensional column-vector with mean $\bs{0}$ and variance $\bs{\mathcal{V}}(\bs{\lambda})$;  $\bs{\mathcal{V}}(\bs{\lambda})=\begin{bmatrix}
	\text{Re}(\bs{\Lambda}^{k}(\bs{\lambda},\bs{\lambda})) & - \text{Im}(\bs{\Lambda}^{k}(\bs{\lambda},\bs{\lambda})) \\
	\text{Im}(\bs{\Lambda}^{k}(\bs{\lambda},\bs{\lambda})) & \text{Re}(\bs{\Lambda}^{k}(\bs{\lambda},\bs{\lambda}))
\end{bmatrix}$ and $\bs{\Lambda}^{k}(\bs{\lambda},\bs{\mu}) = \left[ \Lambda_{(\bs{c},\bs{b})}^{k}(\bs{\lambda},\bs{\mu}) \right]_{(\bs{c},\bs{b})\in[\upsigma_{k}\left(\{1,\dots,d\}\right)]^{2}  }$ with
\begin{align*}
	\hspace{-1.0cm}
	\lim\limits_{T\to\infty}H_{T}^{k-1}T\Cov\left( \hat{g}_{\bs{c},k}^{\Hat{u},(T)}(\bs{\lambda}), \hat{g}_{\bs{b},k}^{\Hat{u},(T)}(\bs{\mu}) \right) &= \Lambda_{(\bs{c},\bs{b})}^{k}(\bs{\lambda},\bs{\mu})\\
	\hspace{-1.0cm}
	{} & \hspace*{-3.5cm} =2\pi \sum_{\upsigma\left(\{1,\dots,k\}\right)}{ \prod_{m=1}^{k}{\eta\left( \lambda_{m} - (\upsigma\mu)_{m} \right)g_{2,(c_{m},(\upsigma{b})_{m})}(\lambda_{m}) }  }\int{W(\bs{\beta})W(\upsigma\bs{\beta})\delta_{D}\left( \sum_{m=1}^{k}{\beta_{m}} \right)d\bs{\beta}}
\end{align*}
where $\upsigma\bs{b}=(b_{\upsigma_{1}},\dots,b_{\upsigma_{k}})$, $\upsigma=(\upsigma_{1},\dots,\upsigma_{k})\in\upsigma\left(\{1,\dots,k\}\right)$, $\sum_{m=1}^{k}{\lambda_{m}}=0[\text{mod}(2\pi)]$, $\eta\left({x}\right)=\sum_{j=-\infty}^{\infty}{\delta_{D}(x+2\pi{j})}$ and $\delta_{D}(x)$ is the delta-Dirac function.

The estimator of our matrix of interest  is $\widehat{\bs{\Pi}}=\text{Re}\left\{\hat{G}_{k}^{\Hat{u},2,(T)}(\bs{\lambda})\right\}$. Besides, given that $\hat{G}_{k}^{\Hat{u},(T)}(\bs{\lambda}) = \vect^{-1}\left( \hat{g}_{k}^{\Hat{u},(T)}(\bs{\lambda}) \right)$,  $\widehat{\bs{\Pi}}$ is a function of real and imaginary parts of $\hat{g}_{k}^{\Hat{u},(T)}(\bs{\lambda})$ as the following equation shows
\begin{align*}
	\hspace{-1.0cm}
	\vect\left(\widehat{\bs{\Pi}}\right)=\vect\left(\text{Re}\left\{\hat{G}_{k}^{\Hat{u},2,(T)}(\bs{\lambda})\right\} \right) &= \left[ \left( I_{d}\otimes{\hat{g}_{\text{Re},k}^{\Hat{u},(T)}(\bs{\lambda})} \right)^{\otimes{2}}  +  \left(I_{d}\otimes{\hat{g}_{\text{Im},k}^{\Hat{u},(T)}(\bs{\lambda})}\right)^{\otimes{2}} \right]\vect(\mb{I}(d))\;,
\end{align*}
where $\hat{g}_{\text{Re},k}^{\Hat{u},(T)}(\bs{\lambda})$ and $\hat{g}_{\text{Im},k}^{\Hat{u},(T)}(\bs{\lambda})$ denote the real and imaginary parts of vector $\hat{g}_{k}^{\Hat{u},(T)}(\bs{\lambda})$, respectively; and $\mb{I}(d)=\vect(\bs{I}_{d})\vect(\bs{I}_{d})^{\prime}\otimes\bs{I}_{d^{k-1}}$.

Consequently, the asymptotic distribution of $\vect\left(\widehat{\bs{\Pi}}\right)$ depends on the null hypothesis at step $s$, $H_{0,s}$. Particularly, at the first step $H_{0,1}:\,\rank(\bs{\Pi})=0$, i.e., under joint Gaussianity of the structural shocks, the population spectrum of order $k$ is zero for $k=3,4$. This result implies that the Jacobian of $\vect\left(\widehat{\bs{\Pi}}\right)$ at population values is null, making it not feasible to use the standard Delta method for finding the asymptotic distribution of $\vect\left(\widehat{\bs{\Pi}}\right)$. Proposition (\ref{prop:Proposition40}) states the asymptotic distribution for  $\vect\left(\widehat{\bs{\Pi}}\right)$ at different stages in the sequential testing procedure.

\begin{proposition} \label{prop:Proposition40}
	Assuming that result in (\ref{eq:AsympSpectEstimator}) holds, then:
	\begin{enumerate}
		\item When $s=1$, under the null hypothesis $H_{0,1}$:
		\begin{align}
			\label{eq:DeltaMethod_Gaussian_G2Sq}
			H_{T}^{k-1}T\left( \vect\left( \widehat{\bs{\Pi}} \right)  \right) &\xrightarrow{d} \bs{Q}\;,
		\end{align}
		
		\item For $s\geq{2}$, under the null hypotesis $H_{0,s}$:
		\begin{align}
			\hspace{-1.0cm}
			\label{eq:DeltaMethod_G2Sq}
			\sqrt{H_{T}^{k-1}T}\left( \vect\left(\widehat{\bs{\Pi}} \right) - \vect\left(\bs{\Pi} \right) \right) &\xrightarrow{d} \mathcal{N}_{d^{2}}\left( \bs{0},\; \bs{J}^{\prime}_{\hat{G}^{2}_{k}}(\bs{\lambda})\bs{\mathcal{V}}(\bs{\lambda})\bs{J}_{\hat{G}^{2}_{k}}(\bs{\lambda}) \right),
		\end{align}
	\end{enumerate}
\end{proposition}
In Proposition \ref{prop:Proposition40}, $\bs{Q}\defeq\frac{1}{2}\left(I_{d^{2}}\otimes{\mb{X}^{\prime}}\right)\bs{\mathcal{H}}_{\hat{G}^{\Hat{u},2,(T)}_{k}}$  and $\bs{\mathcal{H}}_{\hat{G}^{\Hat{u},2,(T)}_{k}}$ is the vectorized Hessian\footnote{The Hessian of a vector-valued function $\mbb{R}^{d}\mapsto{\mbb{R}^{q}}$ is a $3$-tensor. Intuitively, the Hessian matrix is $3$-dimensional array of dimensions $d\times{d}\times{q}$.} of $\vect\left( \text{Re}\left\{\hat{G}_{k}^{\Hat{u},2,(T)}(\lambda)\right\} \right)$ evaluated at zero. $\mb{X}\defeq\vect\left[{\mathcal{W}_{1}(\bs{\mathcal{V}}(\bs{\lambda}))}\right]$ with ${\mathcal{W}_{1}(\bs{\mathcal{V}}(\bs{\lambda}))}$ denotes a Wishart distribution with parameters $\bs{\mathcal{V}}(\bs{\lambda})$ and $1$ degree of freedom. Besides, $\bs{J}_{\hat{G}^{2}_{k}}(\bs{\lambda})$ is the Jacobian of $\vect\left( \text{Re}\left\{\hat{G}_{k}^{2,(T)}(\lambda)\right\} \right)$ evaluated at population values of cumulant spectrum of order $k$.

\subsection{Asymptotic Equivalence between spectrum estimators of ${\bs{u}}_{t}$ and $\hat{\bs{u}}_{t}$}
The results above are obtained for unobserved RF errors, $\bs{u}_{t}$. However, we employ the estimated RF residuals using a finite sample of size $T$ using the estimated RF parameters, $\hat{\bs{\vartheta}}_{f}$. In consequence, it is essential to show that asymptotic results remain valid when using estimated RF residuals, $\hat{\bs{u}}_{t}$. The spectrum of order $k$ for the estimated RF residuals can be obtained easily replacing $\bs{u}_{t}$ by $\hat{\bs{u}}_{t}$ in (\ref{eq:SpectrumK}), denoted by $\hat{g}_{\bs{c},k}^{\Hat{u},(T)}(\bs{\lambda})$. The rest of the estimators can be obtained as explained above using $\hat{g}_{\bs{c},k}^{\Hat{u},(T)}(\bs{\lambda})$. This analysis is omitted in \citet{maxand2020identification} and \citet{guay2021identification} works, maybe explained because the structural linear model is fundamental.

\begin{proposition} \label{prop:Proposition41}
	Given our model is determined by (\ref{eq:Eq1})-(\ref{eq:Eq2}) and under Assumption \ref{as:Assumption1}, it holds 
	\begin{equation*}
		\mbb{E}\sqrt{H_{T}^{k-1}T}\left| \hat{g}_{\bs{c},k}^{u,(T)}(\bs{\lambda}) - \hat{g}_{\bs{c},k}^{\Hat{u},(T)}(\bs{\lambda})\right| \leq C H_{T}^{\frac{k-1}{2}}
	\end{equation*}
\end{proposition}
Proposition \ref{prop:Proposition41} states the asymptotic equivalence between estimators of higher-order cumulant spectrum using unobserved RF errors and their sample counterparts. This implies that the discrepancy generated by using $\hat{g}_{\bs{c},k}^{\Hat{u},(T)}(\bs{\lambda})$ decreases to zero at the rate of $C H_{T}^{\frac{k-1}{2}}$, which is slower than $\sqrt{T}$. This convergence rate comes from the fact that we are using a non-parametric estimation for the higher order spectrum, which incorporates a kernel with band-with $H_{T}$.

\begin{corollary}\mbox{}\newline
	\label{prop:Corollary41}
	Let structural model be described by (\ref{eq:Eq1}) but condition in (\ref{eq:Eq2}) holds for any point in $\mbb{T}_{+}$, then
	\begin{align*}
		\mbb{E}\sqrt{H_{T}^{k-1}T}\left| \hat{g}_{\bs{c},k}^{u,(T)}(\bs{\lambda}) - \hat{g}_{\bs{c},k}^{\Hat{u},(T)}(\bs{\lambda})\right| &\leq C H_{T}^{\frac{k-1}{2}}; \\
		\mbb{E}\sqrt{T}\left| \hat{\bs{\kappa}}^{\bs{u},(T)}_{k} - \hat{\bs{\kappa}}^{\hat{\bs{u}},(T)}_{k}\right| &\leq C {T}^{-\frac{1}{2}}.
	\end{align*}
\end{corollary}
In Corollary \ref{prop:Corollary41}, $\hat{\bs{\kappa}}^{\bs{u},(T)}_{k}$ and $\hat{\bs{\kappa}}^{\hat{\bs{u}},(T)}_{k}$ are the sample estimators for cumulant of order $k$ based on unobserved RF errors and RF residuals, respectively. 

This corollary states that in the case of working with a more restricted structural model, since the researcher imposes the location of roots, Proposition (\ref{prop:Proposition41}) still holds. The asymptotic equivalence remains if contemporaneous higher-order cumulants were employed instead of the higher-order spectrum. However, the convergence rate is faster than in the general setting. This latter result is because the estimation does not use kernel smoothing when using contemporaneous higher-order cumulants.
				
	\section{Simulation Evidence and Empirical Application}
    Show the validity of the restricted bootstrap sampling analytically is quite intricate. We believe that following procedures that have been proven to be consistent may assure the effectiveness of our approach, although this is not equivalent to formal proof. Additionally, we present some evidence from different Montecarlo exercises as additional support for the validity of our bootstrap sampling and test. 

The data-generating process or the true structural model for all the exercises is a non-causal SVAR($1$). We employ different distributions for the structural disturbances. Montecarlo and bootstrap repetitions are set to $m=250$ and $B=500$, respectively. The nominal significance level is set to $\alpha=5\%$. The number of points for the DFT is set to a minimum even integer greater than or equal to the sample size $T$, and the size of the window is set to $\floor{T/4}$.

\subsection{Simulation Results}
Table \ref{tab:Table123_SVARMA10_2D} shows the rejection rates of our sequential procedure in the bivariate case ($d=2$). Besides, the values in a box represent the size of the test. At panel (\ref{tab:Table1_SVARMA10_2D}), we employ the zero frequency ($\lambda=0$). The first row is associated with the Gaussian case. Its size is $4\%$. When we consider a mixed case (second row of the panel (\ref{tab:Table1_SVARMA10_2D})), the rejection rate of $H_{0,1}$ represents the power, $32\%$; while its size is $7\%$. Finally, when structural shocks are fully non-Gaussian distributed (third row in panel (\ref{tab:Table1_SVARMA10_2D})), both rejection rates of $H_{0,1}$ and $H_{0,2}$ represent the power at each step. These values are $42\%$ and $26\%$, respectively.

\begin{table}[!htb]
	\centering
	\caption{Estimating $\rank\text{Re}\left( G_{3}^{2}(\lambda_{3}) \right)$: Rejection Rates (Level: $\alpha=5\%$, Sample: $T=250$)}
	\label{tab:Table123_SVARMA10_2D}
	\begin{minipage}[c][0.275\textheight][t]{0.5\textwidth}
			\begin{center}
			\subcaption{Single Frequency ($\lambda_{3}=0$)}
			\label{tab:Table1_SVARMA10_2D}
			\begin{tabular}{ccc}
				\bottomrule
				Distribution of $\bs{\varepsilon}_{t}$ & $H_{0,0}$ & $H_{0,1}$ \\
				\midrule
				${\mathcal{N}(\bs{0},\bs{I}_{2})}$ & $0.04$ & $0.01$ \\
				${\left(\mathcal{N}(0,1);\;\chi^{2}_{2}\right)}$ & $0.32$ & $0.08$ \\
				${\left(E(0,1);\;\chi^{2}_{2}\right)}$ & $0.42$ & $0.26$ \\
				\bottomrule
			\end{tabular}
			\par\end{center}

			\begin{center}
				\subcaption{Grid of $26$ frequencies}
				\label{tab:Table3_SVARMA10_2D}
				\begin{tabular}{ccc}
					\bottomrule
					Distribution of $\bs{\varepsilon}_{t}$ & $H_{0,0}$ & $H_{0,1}$ \\
					\midrule
					${\mathcal{N}(\bs{0},\bs{I}_{2})}$ & $0.05$ & $0.01$ \\
					${\left(\mathcal{N}(0,1);\;\chi^{2}_{2}\right)}$ & $0.37$ & $0.10$ \\
					${\left(E(0,1);\;\chi^{2}_{2}\right)}$ & $0.53$ & $0.35$ \\
					\bottomrule
				\end{tabular}
			\end{center}
	\end{minipage}\hfill{}%
	\begin{minipage}[c][0.275\textheight][t]{0.5\textwidth}
			\begin{center}
				\subcaption{Grid of $11$ frequencies}
				\label{tab:Table2_SVARMA10_2D}
				\begin{tabular}{ccc}
					\bottomrule
					Distribution of $\bs{\varepsilon}_{t}$ & $H_{0,0}$ & $H_{0,1}$ \\
					\midrule
					${\mathcal{N}(\bs{0},\bs{I}_{2})}$ & $0.04$ & $0.01$ \\
					${\left(\mathcal{N}(0,1);\;\chi^{2}_{2}\right)}$ & $0.33$ & $0.09$ \\
					${\left(E(0,1);\;\chi^{2}_{2}\right)}$ & $0.42$ & $0.27$ \\
					\bottomrule
				\end{tabular}
			\end{center}
	\end{minipage}
\end{table}

Employ estimates at a single frequency of higher order cumulant spectrum may be noisy, leading to imprecise results, especially when the sample size is small. To make our procedure robust to this unpleasant characteristic, we use a grid of frequencies instead of only the zero frequency. Let $\mathcal{G}$ denote the finite grid of frequencies that are selected, instead of constructing our matrix of interest as $\bs{\Pi}=\sum_{{\lambda}_{k}\in\mathcal{F}}{\text{Re}\left(G_{k}^{2}({\lambda_{k}})\right)}$ where $\mathcal{F}$ denotes a finite grid of frequencies, because it may entail the risk of gaining rank spuriously; we use the following statistic $\overline{KP}_{r}^{(T)}=\max_{\lambda\in\mathcal{F}}{\left\{ KP_{r}^{(T)}(\lambda) \right\}}$, where $KP_{r}^{(T)}(\lambda)$ denotes the KP-statistic for a particular frequency $\lambda$. 

In panel (\ref{tab:Table2_SVARMA10_2D}) of Table \ref{tab:Table123_SVARMA10_2D}, a grid of length $11$ is selected. For this grid, the test size when all disturbances are Gaussian distributed is $4\%$; when only one non-Gaussian shock is present, the size is around $6\%$. The power is quite similar to when a single frequency is employed. In panel (\ref{tab:Table3_SVARMA10_2D}) of Table \ref{tab:Table123_SVARMA10_2D}, a grid of $26$ frequencies was selected. The size test when $\bs{\varepsilon}_{t}$ is a Gaussian process is $5\%$; and when we have only non-Gaussian shock, the size is around $6\%$.

Regarding power, the levels remain close to the previous exercises. Furthermore, in Table \ref{tab:Table123_SVARMA10_2D_T500}, we can observe the effect of increasing the sample size. Regarding the test size for each case, the results show that it remains close to the values obtained with $T=250$. Moreover, concerning the power, we observe a significant increment in the power, especially in the case where only one single frequency is employed. 

\begin{table}[!htb]
	\centering
	\caption{Estimating $\rank\left( G_{3}^{2}(\lambda_{3}) \right)$: Rejection Rates (Level: $\alpha=5\%$, Sample: $T=500$)}
	\label{tab:Table123_SVARMA10_2D_T500}
	\begin{minipage}[c][0.275\textheight][t]{0.5\textwidth}
			\begin{center}
			\subcaption{Single frequency ($\lambda_{3}=0$)}
			\label{tab:Table1_SVARMA10_2D_T500}
			\begin{tabular}{ccc}
				\bottomrule
				Distribution of $\bs{\varepsilon}_{t}$ & $H_{0,0}$ & $H_{0,1}$ \\
				\midrule
				${\mathcal{N}(\bs{0},\bs{I}_{2})}$ & $0.04$ & $0.01$ \\
				${\left(\mathcal{N}(0,1);\;\chi^{2}_{2}\right)}$ & $0.44$ & $0.09$ \\
				${\left(E(0,1);\;\chi^{2}_{2}\right)}$ & $0.59$ & $0.31$ \\
				\bottomrule
			\end{tabular}
			\par\end{center}

			\begin{center}
				\subcaption{Grid of $26$ frequencies}
				\label{tab:Table3_SVARMA10_2D_T500}
				\begin{tabular}{ccc}
					\bottomrule
					Distribution of $\bs{\varepsilon}_{t}$ & $H_{0,0}$ & $H_{0,1}$ \\
					\midrule
					${\mathcal{N}(\bs{0},\bs{I}_{2})}$ & $0.03$ & $0.00$ \\
					${\left(\mathcal{N}(0,1);\;\chi^{2}_{2}\right)}$ & $0.55$ & $0.15$ \\
					${\left(E(0,1);\;\chi^{2}_{2}\right)}$ & $0.71$ & $0.51$ \\
					\bottomrule
				\end{tabular}
			\end{center}
	\end{minipage}\hfill{}%
	\begin{minipage}[c][0.275\textheight][t]{0.5\textwidth}
			\begin{center}
				\subcaption{Grid of $11$ frequencies}
				\label{tab:Table2_SVARMA10_2D_T500}
				\begin{tabular}{ccc}
					\bottomrule
					Distribution of $\bs{\varepsilon}_{t}$ & $H_{0,0}$ & $H_{0,1}$ \\
					\midrule
					${\mathcal{N}(\bs{0},\bs{I}_{2})}$ & $0.03$ & $0.00$ \\
					${\left(\mathcal{N}(0,1);\;\chi^{2}_{2}\right)}$ & $0.46$ & $0.11$ \\
					${\left(E(0,1);\;\chi^{2}_{2}\right)}$ & $0.62$ & $0.37$ \\
					\bottomrule
				\end{tabular}
			\end{center}
	\end{minipage}
\end{table}

Suppose we take a strategy for restricting the bootstrap sample as the one followed by \citet{guay2021identification}, which employs the first $r_{s}$ columns of $\bs{R}_{1}$ for projecting the residuals into the non-Gaussian dimension. In our case, since our matrix of interest is squared and symmetric, $\bs{R}_{1}=\bs{R}_{2}$, thus we use the first $r_{s}$ columns of $\bs{R}_{2}$. The rest of the components $d-r_{s}$ are drawn from a multivariate normal distribution with mean zero and variance $\bs{I}_{d-r_{s}}$. In Table (\ref{tab:Table123_SVARMA10_2D_Guay}), we compute the rejection rates using Guay's bootstrap sampling approach. The size of the test when having fully Gaussian shocks is, disregarding the length of the frequency grid, around $5\%$. The size when $\bs{\varepsilon}_{t}$ has one Gaussian component is around $7\%$ for the different grid lengths. Nonetheless, the power using \citet{nordhausen2017asymptotic} bootstrap sampling is consistently higher than using Guay's approach.

\begin{table}[!htb]
	\centering
	\caption{Estimating $\rank\left( G_{3}^{2}(\lambda_{3}) \right)$: Rejection Rates (Level: $\alpha=5\%$, Sample: $T=250$) \\ (Bootstrap sample constructed following \citet{guay2021identification})}
	\label{tab:Table123_SVARMA10_2D_Guay}
	\begin{minipage}[c][0.275\textheight][t]{0.5\textwidth}
			\begin{center}
			\subcaption{Single frequency ($\lambda_{3}=0$)}
			\label{tab:Table1_SVARMA10_2D_Guay}
			\begin{tabular}{ccc}
				\bottomrule
				Distribution of $\bs{\varepsilon}_{t}$ & $H_{0,0}$ & $H_{0,1}$ \\
				\midrule
				${\mathcal{N}(\bs{0},\bs{I}_{2})}$ & $0.04$ & $0.01$ \\
				${\left(\mathcal{N}(0,1);\;\chi^{2}_{2}\right)}$ & $0.24$ & $0.07$ \\
				${\left(E(0,1);\;\chi^{2}_{2}\right)}$ & $0.40$ & $0.23$ \\
				\bottomrule
			\end{tabular}
			\par\end{center}

			\begin{center}
				\subcaption{Grid of $26$ frequencies}
				\label{tab:Table3_SVARMA10_2D_Guay}
				\begin{tabular}{ccc}
					\bottomrule
					Distribution of $\bs{\varepsilon}_{t}$ & $H_{0,0}$ & $H_{0,1}$ \\
					\midrule
					${\mathcal{N}(\bs{0},\bs{I}_{2})}$ & $0.05$ & $0.01$ \\
					${\left(\mathcal{N}(0,1);\;\chi^{2}_{2}\right)}$ & $0.24$ & $0.07$ \\
					${\left(E(0,1);\;\chi^{2}_{2}\right)}$ & $0.42$ & $0.27$ \\
					\bottomrule
				\end{tabular}
			\end{center}
	\end{minipage}\hfill{}%
	\begin{minipage}[c][0.275\textheight][t]{0.5\textwidth}
			\begin{center}
				\subcaption{Grid of $11$ frequencies}
				\label{tab:Table2_SVARMA10_2D_Guay}
				\begin{tabular}{ccc}
					\bottomrule
					Distribution of $\bs{\varepsilon}_{t}$ & $H_{0,0}$ & $H_{0,1}$ \\
					\midrule
					${\mathcal{N}(\bs{0},\bs{I}_{2})}$ & $0.04$ & $0.01$ \\
					${\left(\mathcal{N}(0,1);\;\chi^{2}_{2}\right)}$ & $0.21$ & $0.07$ \\
					${\left(E(0,1);\;\chi^{2}_{2}\right)}$ & $0.39$ & $0.24$ \\
					\bottomrule
				\end{tabular}
			\end{center}
	\end{minipage}
\end{table}

For larger dimensional models, we choose $d=3,4$. We perform these exercises with sample size $T=250$ and the grid $\lambda_{3}=0$. The results can be observed in Table (\ref{tab:Table123_SVARMA10_4D}). The distribution $MN_{1}$ is a mixture of two normal distributions ($\mathcal{N}(10,0.75)$ and $\mathcal{N}(-2,4)$) such that the distribution has a positive skewness coefficient. It can be noted that, while the sample size is fixed, it is more difficult to reject the null hypothesis for larger $r_{s}$.

\begin{table}[!htb]
	\centering
	\caption{Estimating $\rank\left( G_{3}^{2}(\lambda_{3}) \right)$: Rejection Rates (Level: $\alpha=5\%$, Sample: $T=250$)}
	\label{tab:Table123_SVARMA10_4D}
	\begin{minipage}[c][0.1\textheight][t]{0.75\textwidth}
			\begin{center}
			\label{tab:Table1_SVARMA10_4D}
			\begin{tabular}{ccccc}
				\bottomrule
				Distribution of $\bs{\varepsilon}_{t}$ & $H_{0,0}$ & $H_{0,1}$ & $H_{0,2}$ & $H_{0,3}$ \\
				\midrule
				${\mathcal{N}(\bs{0},\bs{I}_{3})}$ & $0.03$ & $0.02$ & $0.02$ & $--$ \\
				${\left(E(0,1);\;MN_{1};\;\mathcal{N}(\bs{0},1)\right)}$ & $0.36$ & $0.21$ & $0.08$ & $--$ \\
				${\left(\mathcal{N}(\bs{0},\bs{I}_{2});\;E(0,1);\;\chi^{2}_{2}\right)}$ & $0.57$ & $0.27$ & $0.04$ & $0.01$ \\
				\bottomrule
			\end{tabular}
			\par\end{center}
	\end{minipage}
\end{table}
\vspace*{-0.5cm}
\subsection{Empirical Application}
\label{sec:EmpiricApp}
We select two well-known data sets in empirical macroeconomics. The first data set is used in the seminal work of \citeauthor{blanchard1989dynamic} (hereafter BQ), where they introduced the long-run restrictions as an identification scheme for causal structural VAR models. This database contains two endogenous variables: the US GNP's growth and the unemployment rate. The second data set is taken from the work of \citet{blanchard2002empirical} (hereafter BP). Their data includes three endogenous variables: tax revenues, government spending, and GDP (all in real terms). In both cases, we use the exact SVAR specification for each work, avoiding our conclusions from being affected by a specification bias.

Table \ref{tab:ResultApplication_BQ} shows the results of applying our proposal to the BQ database. We can identify a single asymmetric structural shock using only third-order cumulant spectral density. When information in the fourth-order spectral cumulant is employed, the proposed method cannot identify any component with excess kurtosis. This result can seem contradictory, though the limited sample size ($T=148$) might affect the precision of fourth-order estimates. On the other hand, Table \ref{tab:ResultApplication_BP} shows the results of applying our proposed method to the BP dataset. Using third-order information, we detect at most two asymmetric structural shocks at a significance level of $10\%$. This result is consistently found disregarding if only a single frequency or a grid of frequencies is employed. When fourth-order information is used, at $10\%$ of significance, we can detect three non-mesokurtic structural shocks at zero frequency. When a grid of frequencies is employed instead, the method does detect two non-mesokurtic structural shocks. Besides, if these results are compared to those obtained in \citet{guay2021identification}, our proposal can detect at least one extra non-mesokurtic structural shock.
\newpage
\begin{table}[!tbh]
	\centering
	\caption{Number of non-Gaussian components in \citet{blanchard1989dynamic}}
	\label{tab:ResultApplication_BQ}
	\begin{minipage}[c][0.25\textheight][t]{0.5\textwidth}
		\begin{center}
			\subcaption{$\rank\left( \text{Re}\left({G}_{3}^{2}(\lambda)\right) \right)$}
			\begin{tabular}{lccc}
				\\[-1.8ex]\toprule 
				\toprule \\[-1.8ex]
				Frequency & & \multicolumn{2}{c}{Null Hypotheses} \\
				\cmidrule{3-4} Grid & & $H_{0,1}$ & $H_{0,2}$ \\
				[0.5ex]\midrule \\[-1.8ex]
				& & & \\
				\multirow{1}{*}{$\lambda=0$} & & $0.077$ & $0.160$ \\
				\multirow{1}{*}{Grid of $11$ freqs.} & & $0.037$ &  $0.128$ \\
				\multirow{1}{*}{Grid of $26$ freqs.} & & $0.093$ &  $0.157$ \\
				\bottomrule
				\bottomrule \\[-1.8ex]
				\multicolumn{4}{l}{\scriptsize P-values reported, based on $B=1000$ bootstrap samples.}
			\end{tabular}
			\par\end{center}
	\end{minipage}\hfill{}%
	\begin{minipage}[c][0.25\textheight][t]{0.5\textwidth}
		\begin{center}
			\subcaption{$\rank\left( \text{Re}\left({G}_{4}^{2}(\lambda)\right) \right)$}
			\begin{tabular}{lccc}
				\\[-1.8ex]\toprule 
				\toprule \\[-1.8ex]
				Frequency & & \multicolumn{2}{c}{Null Hypotheses} \\
				\cmidrule{3-4} Grid & & $H_{0,1}$ & $H_{0,2}$ \\
				[0.5ex]\midrule \\[-1.8ex]
				& & & \\
				\multirow{1}{*}{$\lambda=0$} & & $0.659$ & $--$ \\
				\multirow{1}{*}{Grid of $11$ freqs.} & & $0.723$ &  $--$ \\
				\multirow{1}{*}{Grid of $26$ freqs.} & & $0.741$ &  $--$ \\
				\bottomrule
				\bottomrule \\[-1.8ex]
				\multicolumn{4}{l}{\scriptsize P-values reported, based on $B=500$ bootstrap samples.}
			\end{tabular}
			\par\end{center}
	\end{minipage}
\end{table}

\begin{table}[!tbh]
	\centering
	\caption{Number of non-Gaussian components in \citet{blanchard2002empirical}}
	\label{tab:ResultApplication_BP}
	\begin{minipage}[c][0.22\textheight][t]{0.5\textwidth}
		\begin{center}
			\subcaption{$\rank\left( \text{Re}\left({G}_{3}^{2}(\lambda)\right) \right)$}
			\resizebox{7.5cm}{!}{
				\begin{tabular}{lcccc}
					\\[-1.8ex]\toprule 
					\toprule \\[-1.8ex]
					Frequency & & \multicolumn{2}{c}{Null Hypotheses} \\
					\cmidrule{3-5} Grid & & $H_{0,1}$ & $H_{0,2}$ & $H_{0,3}$ \\
					[0.5ex]\midrule \\[-1.8ex]
					& & & & \\
					\multirow{1}{*}{$\lambda=0$} & & $0.077$ & $0.021$ & $0.377$ \\
					\multirow{1}{*}{Grid of $11$ freqs.} & & $0.066$ &  $0.041$ & $0.729$ \\
					\multirow{1}{*}{Grid of $26$ freqs.} & & $0.038$ &  $0.064$ & $0.450$ \\
					\bottomrule
					\bottomrule \\[-1.8ex]
					\multicolumn{4}{l}{\scriptsize P-values reported, based on $B=1000$ bootstrap samples.}
				\end{tabular}
			}
			\par\end{center}
	\end{minipage}\hfill{}%
	\begin{minipage}[c][0.22\textheight][t]{0.5\textwidth}		
		\begin{center}
			\subcaption{$\rank\left( \text{Re}\left({G}_{4}^{2}(\lambda)\right) \right)$}
			\resizebox{7.5cm}{!}{
				\begin{tabular}{lcccc}
					\\[-1.8ex]\toprule 
					\toprule \\[-1.8ex]
					Frequency & & \multicolumn{3}{c}{Null Hypotheses} \\
					\cmidrule{3-5} Grid & & $H_{0,1}$ & $H_{0,2}$ & $H_{0,3}$ \\
					[0.5ex]\midrule \\[-1.8ex]
					& & & \\
					\multirow{1}{*}{$\lambda=0$} & & $0.034$ & $0.029$ & $0.068$ \\
					\multirow{1}{*}{Grid of $11$ freqs.} & & $0.042$ &  $0.044$ & $0.103$ \\
					\multirow{1}{*}{Grid of $26$ freqs.} & & $0.024$ &  $0.094$ & $0.252$ \\
					\bottomrule
					\bottomrule \\[-1.8ex]
					\multicolumn{4}{l}{\scriptsize P-values reported, based on $B=250$ bootstrap samples.}
				\end{tabular}
			}
			\par\end{center}
	\end{minipage}
\end{table}
From the previous results, provided that structural shocks are independent across time and components, if the researcher imposes fundamentalness, applying the SIS to both BQ and BP datasets is feasible. A causal and invertible SVARMA model can be identified without imposing external identification restrictions or using proxy variables. Moreover, our results suggest that for the BP dataset, it is feasible to identify a possibly non-fundamental SVARMA model. In the case of the BQ dataset, the most limiting factor is the quite small sample size.
	
	\section{Conclusion and Final Remarks}
	This paper aims to design a procedure to determine the number of non-Gaussian shocks in a structural linear VARMA model which is robust to the type of dynamic representation, i.e., whether the structural model is fundamental or non-fundamental. This objective is mainly motivated because knowing the number of non-Gaussian shocks in the structural model allows the researcher to implement an estimation procedure of structural parameters based on the statistical identification approach. In the fundamentalness of the structural model, the requirement is to have at most one Gaussian structural shock; if the researcher does not want to impose the root location, the requirement is that all the structural errors are non-Gaussian distributed. 
	
	We generalize the procedure in \citet{guay2021identification} by exploiting that the rank of a matrix constructed from the third-order cumulant spectrum of the RF errors reveals the number of skewed or asymmetric structural errors in the SVARMA model. Meanwhile, the rank of an array constructed from the fourth-order cumulant spectrum reveals the number of non-mesokurtic structural shocks in the system. Simulation results show that our procedure correctly estimates the non-Gaussian dimension. Additionally, from a practice point of view, our proposal is intensive computationally, especially if we employ the cumulant spectrum of fourth order or when the dimension of the structural model increases or the sample size is large.
	
	\begin{singlespace}
		\bibliographystyle{apalike}
		\bibliography{references}
	\end{singlespace}
		
\end{document}